\documentclass[12pt,preprint]{aastex}
\usepackage{mathrsfs, amsmath}
\begin{document}

\title{
    Diagnosing Circumstellar Debris Disks\vspace*{2.0in}
}

\author{
    Joseph M. Hahn
}

\affil{
    Space Science Institute\\
    10500 Loring Drive\\
    Austin, TX, 78750\\
    jhahn@spacescience.org\\
    512-291-2255\vspace*{3.0in}
}

\author{
    Review draft version 2\\
    Submitted for publication in the\\
    {\it Astrophysical Journal} on April 21, 2010\\
    Revised June 16, 2010\\
    Accepted June 21, 2010
    \ \vspace*{0.4in}
}

%%%%%%%%%%%%%%%%%%%%%%%%%%%%%%%%%%%
% Abstract.
%%%%%%%%%%%%%%%%%%%%%%%%%%%%%%%%%%%
\begin{abstract}

A numerical model of a circumstellar debris disk is developed and applied to
observations of the circumstellar dust orbiting $\beta$ Pictoris.  The model
accounts for the rates at which dust is produced 
by collisions among unseen planetesimals, and the rate at which
dust grains are destroyed due to collisions. The model also
accounts for the effects of radiation pressure, which is the dominant
perturbation on the disk's smaller but abundant dust grains.
Solving the resulting system of rate equations then provides
the dust abundances versus grain size and over time.
Those solutions also provide the dust grains' collisional 
lifetime versus grain size, and the debris disk's
optical depth and surface brightness versus distance from the star.
Comparison to observations then yields estimates of the unseen planetesimal
disk's radius, and the rate at which the disk sheds mass due to
planetesimal grinding. The model can also be used to measure or else constrain
the dust grain's physical and optical properties, such as the dust grains'
strength, their light scattering 
asymmetry parameter, and the grains' efficiency of light scattering $Q_s$.

The model is then applied to optical observations of the edge-on dust disk
orbiting $\beta$ Pictoris, and good agreement is achieved when the
unseen planetesimal disk is broad, with $75\lesssim r\lesssim150$ AU. If it is assumed
that the dust grains are bright like Saturn's icy rings ($Q_s=0.7$),
then the cross section of dust in the disk is $A_d\simeq2\times10^{20}$ km$^2$ and its
mass is $M_d\simeq11$ lunar masses. In this case the planetesimal  disk's
dust production rate is quite heavy, $\dot{M}_d\sim9$ M$_\oplus$/Myr,
implying that there is or was a substantial amount of planetesimal mass there,
at least 110 earth-masses. But if the dust grains are darker than assumed,
then the planetesimal disk's mass-loss rate and its total mass are heavier.
In fact, the apparent dearth of any major planets in this region,
plus the planetesimal disk's heavy mass-loss rate,
suggests that the $75\lesssim r<150$ AU zone at $\beta$ Pic
might be a region of planetesimal destruction,
rather than a site of ongoing planet formation.

\end{abstract}
%\keywords{key: words\vspace*{1.5in}}

%%%%%%%%%%%%%%%%%%%%%%%%%%%%%%%%%%%
% Introduction.
%%%%%%%%%%%%%%%%%%%%%%%%%%%%%%%%%%%
\section{Introduction}
\label{intro}

A debris disk is a dusty circumstellar disk that is often found 
in orbit about a young star. It is also suspected that these dust disks
might be sites of ongoing planet formation. This thinking is motivated 
by the dust grains' lifetime due to collisions, which is often much
shorter than the age of the host star. Evidently, a circumstellar debris disk is
also being supplied with fresh dust, and collisions among unseen planetesimals
provides a plausible explanation for this dust production.
And because planetesimals are also the seeds of planets,
it is conceivable that a debris disk might be forming planets
as well.

But keep in mind that if the collisional grinding
in the planetesimal disk is too vigorous, then it
is possible that the planetesimals  might instead grind away before
they have a chance to assemble into planets. Indeed, models of the early
evolution of the outer Solar System propose that
the early Kuiper Belt, which is a swarm of comets orbiting beyond Neptune,
was initially composed of $M_{KB}\sim30$ Earth
masses\footnote{The \cite{K02} model
recommends 10 Earth masses in a 6 AU-wide annulus centered at $r=35$ AU,
but this should be multiplied by $\sim3$ to account for the Kuiper Belt's
full radial width.} \citep{K02}, which would be enough to allow for the formation
of two Neptune-class planets. But that model also shows that runaway accretion 
during the next $t\sim500$ Myrs only managed to produce a handful of Pluto-sized
bodies, while the bulk of the planetesimal mass there remained locked in
the smaller planetesimals. Meanwhile, collisions among the smaller bodies
steadily ground much of that mass to dust that
is then blown out of the system by radiation pressure,
which also stalls further growth.
The mass-loss rate implied by this collisional grinding is
$\dot{M}_d\sim t/M_{KB}\sim10^{13}$ gm/sec \citep{SC97, K02}.
Another outer Solar System scenario is the Nice model,
which requires its primordial Kuiper Belt to persist until
$t\sim700$ Myrs since formation,
which is when the giant planets suddenly adjust their orbits
and trigger the Late Heavy Bombardment \citep{GLT05, LMV08}.
Although the dynamical history of the Nice model is rather different from
other models, it should be noted that this and most other models of the
outer Solar System's early evolution predict that the primordial Kuiper Belt
had a quiescent period lasting several hundreds of
millions years, during which the Belt would have lost tens of Earth-masses
of material due to collisional grinding and blowout of dust due to
radiation pressure \citep{SC97, K02}. 

An interesting question is whether the observed
circumstellar debris disks have mass-loss rates
comparable to that predicted for our early Solar System. The answer to that
question will then provide
some guidance as to whether these disks should instead be
thought of potential sites for planet formation,
or else regions of planetesimal destruction.
To address this issue, the following develops a model 
that follows the time-evolution of a circumstellar debris disk.
Sections \ref{orbits}--\ref{evolution} derives in some detail the model's
physics, but readers not interested in those details can skip those sections.
Sections \ref{example}--\ref{relic} examines how the debris disk's
structure and appearance depends upon the system parameters, and those results
can also be quickly gleaned by inspecting the figures there.
Section \ref{beta_pic}
then applies the model to observations of the debris disk orbiting
$\beta$ Pictoris, with Section \ref{summary} providing a summary of
the findings.

%%%%%%%%%%%%%%%%%%%%%%%%%%%%%%%%%%%
% Debris disk model
%%%%%%%%%%%%%%%%%%%%%%%%%%%%%%%%%%%
\section{The debris disk model}
\label{model}

Radiation pressure is the dominate perturbation
on small dust grains orbiting in a circumstellar debris disk
({\it c.f.}, \citealt{SC06}),
and the review in Section \ref{orbits} shows how the resulting dust
orbits are simple functions of grain size. Dust collision rates
are derived in Sections \ref{streamlines}--\ref{collisions},
and Section \ref{evolution} shows how to use those rates to calculate the 
time evolution of the debris disk's dust abundance. The remaining subsections
then illustrate how the simulated debris disk varies with the model parameters.

%%%%%%%%%%%%%%%%%%%%%%%%%%%%%%%%%%%
% Dust Orbit Elements.
%%%%%%%%%%%%%%%%%%%%%%%%%%%%%%%%%%%
\subsection{dust orbit elements}
\label{orbits}

An orbiting dust grain is characterized by its size parameter $\beta$,
which is the ratio of stellar radiation pressure to gravity. For a spherical
grain of radius $R$, the size parameter is 
\begin{equation}
    \label{beta}
    \beta = \frac{3L_\star Q_{rp}}{16\pi GM_\star c\rho R}
        = 0.57Q_{rp}\left(\frac{L_\star}{L_\odot}\right)
               \left(\frac{M_\star}{M_\odot}\right)^{-1}
               \left(\frac{\rho}{\mbox{1 gm/cm$^3$}}\right)^{-1}
               \left(\frac{R}{\mbox{1 $\mu$m}}\right)^{-1}
\end{equation}
where $L_\star$ and $M_\star$ are the star's luminosity and mass, $G$ is
the gravitational constant, $c$ the speed of light, $\rho$ is the grain's volume
density, and $L_\odot$ and $M_\odot$ refer to solar values
\citep{BLS79}. The grain's radiation pressure efficiency is
$Q_{rp}=Q_a + (1-g)Q_s$ \citep{BLS79}, where $Q_a$ is the efficiency
of the grain's absorption of starlight and $Q_s$ is the efficiency
of light scattering by the grain. The scattering asymmetry parameter $g$ is
\begin{equation}
    \label{g}
    g = \int\Phi(\phi)\cos\phi d\Omega
\end{equation}
where the integral runs over all solid angles, and the phase function
$\Phi(\phi)$ gives the proportion of light that is scattered through
scattering angle $\phi$, which is the angle between the directions 
followed by the incident and scattered light. The phase function is normalized
such that $\int\Phi(\phi)d\Omega=1$, so $|g|\le1$. Forward scattering dust
grains have values of $g>0$, backscattering grains have $g<0$, and 
isotropic light scattering has $g=0$. The model developed here will be applied
to the $\beta$ Pictoris debris disk, whose dust grains are substantially
larger than the wavelength of the incident light. This is the geometric
optics limit, and energy conservation in this limit requires $Q_a + Q_s = 1$, so
\begin{equation}
    \label{Q_rp0}
    Q_{rp} = 1-gQ_s.
\end{equation}
Evidently, $Q_{rp}\simeq1$ except when $|gQ_s|$ is not small, which can occur
if the dust grains are efficient forward or back scatters.
Equation (\ref{beta}) shows that 
radiation pressure is significant for grains of radii
$R\sim{\cal O}(1)\ \mu$m when orbiting a solar-type star, and is
unimportant for grains with $R\gg1\ \mu$m. Since radiation pressure has the
same inverse-square force law as stellar gravity, the effects of
radiation pressure are easily accounted for by substituting
$GM_\star\rightarrow(1-\beta)GM_\star$ into all equations for the dust
grains' motion.

Dust is manufactured when planetesimals collide. Consider a dust grain
that is generated by colliding planetesimals that reside in nearly
circular orbits of radius $r$. These planetesimals have a specific energy
$E_p=$ kinetic + potential energy that is
\begin{equation}
    \label{E_p}
    E_p = \frac{1}{2}v^2 - \frac{GM_\star}{r}=-\frac{GM_\star}{2r} 
\end{equation}
where $v=\sqrt{GM_\star/r}$ is the planetesimal's velocity,
assuming its orbit is circular or nearly so. However, a dust grain that
forms via a collision at radius $r$ will have a specific energy 
$E_d = \onehalf v^2 - GM_\star(1-\beta)/r = 
-GM_\star(1-\beta)/2a$, where $a$ is the dust grain's semimajor axis.
This assumes that the dust grain has the same
velocity $v$ as its parent planetesimal at the moment of its
creation, or equivalently, the velocity at which the grain is ejected from
the planetesimal is small compared to its orbital speed.
But this is reasonable since the debris disks considered here are thin, which
implies that dust generation occurs with low ejection speeds.
In this case, $\onehalf v^2=GM_\star/2r$ in the above, so the dust grain's
semimajor axis is \citep{TAB03}
\begin{equation}
    \label{a}
    a(\beta) = \frac{1-\beta}{1-2\beta}r.
\end{equation}
Bound orbits have $a>0$, so only dust having $\beta<\beta_{max}$ where
$\beta_{max}=\onehalf$ will populate the resulting debris disk, while dust having
$\beta\ge\beta_{max}$ are not bound to the star
and quickly leave the system. According to Equation (\ref{beta}),
the smallest bound grains have radii
\begin{equation}
    \label{R_min}
    R_{min} = \frac{3L_\star Q_{rp}}{16\pi\beta_{max} GM_\star c\rho}
        = 1.1Q_{rp}\left(\frac{L_\star}{L_\odot}\right)
               \left(\frac{M_\star}{M_\odot}\right)^{-1}
               \left(\frac{\rho}{\mbox{1 gm/cm$^3$}}\right)^{-1}
               \mbox{ $\mu$m}.
\end{equation}

Next, note that a planetesimal's specific angular momentum $L_p=rv$
also equals the dust grain's angular momentum $L_d$, since both have the same
position and velocity at the moment of the grain's creation. Since a dust
grain's Keplerian orbit has $L_d=\sqrt{GM_\star(1-\beta)a(1-e^2)}$ while the
planetesimal's $L_p=\sqrt{GM_\star r}=L_d$, this provides the grain's eccentricity
\begin{eqnarray}
    \label{e}
    e(\beta) = \frac{\beta}{1-\beta}
\end{eqnarray}
\citep{TAB03}.
The dust grain's periapse $q_{peri}$ and apoapse $Q_{apo}$ distances are then
\begin{mathletters}
    \label{peri/apo}
    \begin{eqnarray}
        q_{peri} &=& a(1-e) = r\\
        \mbox{and}\quad Q_{apo} &=& a(1+e) = r/(1-2\beta)
    \end{eqnarray}
\end{mathletters}
where $r$ is the orbital radii of the planetesimals that gave birth
to the dust. Also
note that periapse is precisely at the planetesimal's orbit, which
tells us that the grain's longitude of periapse $\tilde{\omega}$
is also the longitude where the grain formed.

%%%%%%%%%%%%%%%%%%%%%%%%%%%%%%%%%%%
% streamlines.
%%%%%%%%%%%%%%%%%%%%%%%%%%%%%%%%%%%
\subsection{streamlines}
\label{streamlines}

Assessing the collisions that a dust grain experiences over its 
lifetime can be a laborious calculation, one that usually requires evaluating
integrals over the dust grains' size distribution
and the debris disk's three dimensional volume.
However those integrals can be replaced by a simple summation when
the model disk is quantized; see for example  \cite{KLS06}. In the following, 
the planetesimal disk, which is the source of this dust,
is represented by $N_r$ concentric rings that have
radii $r$ uniformly distributed
over the interval $r_{in}\le r\le r_{out}$. Each planetesimal ring also
produces dust at $N_l$ discreet sites that are distributed
uniformly in longitude about the ring.
The dust grains' radii $R=(\beta_{max}/\beta)R_{min}$ are also quantized so
that they uniformly sample the interval $R_{min}<R<R_{max}$,
with $N_\beta$ size-bins in the interval $0<\beta<\beta_{max}$. 
As a result, the planetesimal
disk produces dust whose grains inhabit $N_s=N_rN_lN_\beta$ distinct streamlines,
or orbits, whose shapes are given by Equations (\ref{a}) and (\ref{e}),
and whose orientations are uniformly distributed in longitude.

The rate of dust production $p(R, r)$ by the planetesimal disk
is also a power law, with $p(R, r)\propto R^{-q}$ being the rate at which
the planetesimals inject dust of radii $R$ into a single streamline, 
with $q$ being the power law for the grains' differential size distribution.
The model will also allow for the dust production rate to vary
with the source planetesimals' radial distance $r$ as
$p(R, r)\propto(r/r_{out})^c$. A planetesimal disk having $c<0$
can then be thought of as experiencing `inside-out' erosion, since the
inner portion of that disk experiences greater dust production than its
outer parts, while a disk having $c>0$ suffers `outside-in' erosion.
Also note that $R\propto\beta^{-1}$, so the dust production rate
can be written $p(\beta, r)=p_0(\beta/\beta_{max})^{q}(r/r_{out})^c$
where $p_0$ is a constant.

Since there are $N_s=N_rN_lN_\beta$ streamlines in the model debris disk,
the index $i$ will be used to identify any one such dusty streamline, 
where $1\le i\le N_s$.  The streamlines' longitudes of periapse $\tilde{\omega}_i$
are uniformly distributed such that the resulting debris disk
has $N_l$--fold symmetry. The quantity $n_i(t)$ will be the number of dust
grains inhabiting streamline $i$ at time $t$; these grains all have
the same size $\beta_i$ and orbit elements $a_i, e_i, \tilde{\omega}_i$,
and they all formed at the same longitude $\tilde{\omega}_i$
in the same planetesimal ring that has radius $r_i$.
The streamline's abundance $n_i(t)$ will then
evolve over time $t$ due to the production of dust
by the planetesimal ring $r_i$, but that quantity will also vary
due to collisions with the dust grains
that inhabit the disk's other streamlines. The rate at which these dust grains
collide and destroy each other is derived below.

%Fig: collision sketch
\begin{figure}[t]
\epsscale{0.75}
\plotone{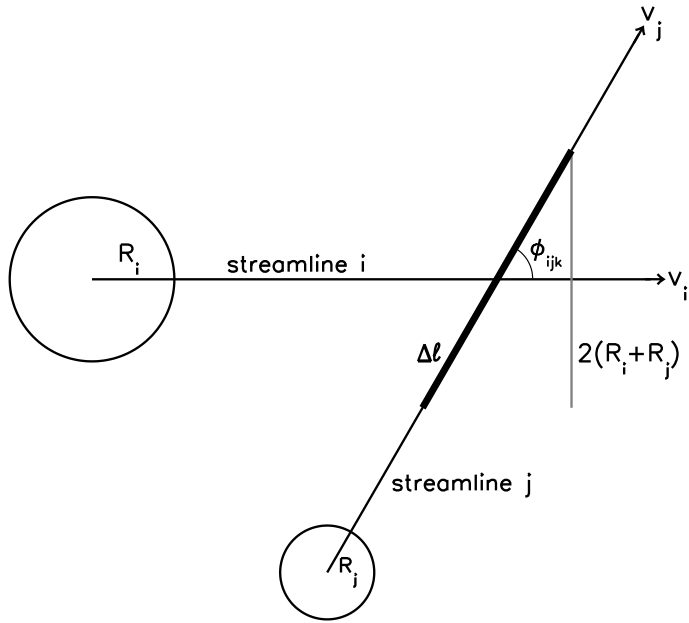}
\figcaption{
    \label{collision_fig}
    A dust grain of radius $R_i$ crosses streamline $j$ that contains dust
    of radii $R_j$. Grain $i$ sweeps
    a length $\Delta\ell=2(R_i+R_j)/\sin\phi_{ijk}$ along streamline $j$ where
    $\sin\phi_{ijk}$ is the angle between the grains' velocity vectors
    $\mathbf{v}_i$ and $\mathbf{v}_j$.
}
\end{figure}

%%%%%%%%%%%%%%%%%%%%%%%%%%%%%%%%%%%
% collisions
%%%%%%%%%%%%%%%%%%%%%%%%%%%%%%%%%%%
\subsection{collision rates}
\label{collisions}

Figure \ref{collision_fig} shows a dust grain in streamline $i$ as it crosses
streamline $j$. The dust in these streamlines have radii $R_i$ and $R_j$
and velocities $\mathbf{v}_i$ and $\mathbf{v}_j$. Suppose for now that the
debris disk is flat, and that all dust grains have zero inclinations.
In this case, grain $i$ would sweep across a length
$\Delta\ell=2(R_i+R_j)/\sin\phi_{ijk}$ in streamline $j$,
where $\phi_{ijk}$ is the angle between vectors
$\mathbf{v}_i$ and $\mathbf{v}_j$, so
$\sin\phi_{ijk}=|\mathbf{v}_i\times\mathbf{v}_j|/v_iv_j$.
This quantity depends not just on the streamlines' orbit elements
$a_i, e_i$  and $a_j, e_j$, but also
on the streamlines' relative longitude of periapse
$\tilde{\omega}_k = \tilde{\omega}_j - \tilde{\omega}_i$, which is why
the $k$ subscript is also introduced.

The linear density of dust grains in streamline $j$ is $\lambda_j$,
so particle $i$ will collide with $\lambda_j\Delta\ell$ particles
as it traverses streamline $j$. The
flux of dust in this streamline is $\lambda_j v_j$, which is the rate at which grains
cross a point in that streamline, so
$\lambda_j v_j=n_j/T_j$ where $n_j$ is the total number of grains in that streamline
and $T_j=2\pi\sqrt{a_j^3/(1-\beta_j)GM_\star}$ is that streamline's orbit period.
So the number of collisions that grain $i$ suffers as it crosses this site in
streamline $j$ is $\Delta n_{ijk}=\lambda_j\Delta\ell = 
2(R_i+R_j)n_jf_z/\sin\phi_{ijk}v_jT_j$, where the factor
$f_z$ is now introduced to account for the dust disk's vertical thickness.
Note that $f_z$ would be unity if streamlines in the dust-disk where coplanar.
However a real disk has a vertical half-width $h=Ir_{ijk}$,
where $I$ is the grains' characteristic inclination and $r_{ijk}$ is the
radial distance from the star where streamlines $i$ and $j$ cross,
with the $k$ subscript again indicating that
that this quantity also depends on the streamlines'
relative longitude of periapse $\tilde{\omega}_k$.
Thus $f_z=2f_c(R_i + R_j)/2h=f_c(R_i + R_j)/Ir_{ijk}$ 
is the probability that two grains in
streamlines $i$ and $j$ are close enough in the vertical direction to come in
contact, and the factor $f_c=\pi/4$ is a geometric correction that accounts
for the grain's circular cross-section. And since there are $n_i$
grains in streamline $i$, they will suffer $n_i\Delta n_{ijk}$ collisions
with streamline $j$
after one orbit period $T_i$, so the grains in streamline $i$ collide with
the grains in streamline $j$ at site $r_{ijk}$ at the rate
$n_i\Delta n_{ijk}/T_i$.
Thus the total rate at which grains in streamline $i$ suffer
collisions with all of the disk's other streamlines is
\begin{eqnarray}
    \label{dN^c}
    {\cal R}_i = \sum\frac{n_i\Delta n_{ijk}}{T_i}
        =\frac{n_i}{T_{out}}\sum_{j=1}^{N_rN_\beta}\sum_{k=1}^{N_l}
        \sum_{r_1}^{r_2}\alpha_{ijk}n_j,
\end{eqnarray}
where the coefficient
$\alpha_{ijk}=\pi(R_i+R_j)^2T_{out}f^c_{ijk}/2I\sin\phi_{ijk}r_{ijk}v_jT_iT_j$,
and $T_{out}=2\pi\sqrt{r_{out}^3/GM_\star}$
is the orbit period of the outermost planetesimal ring at $r=r_{out}$.
Note that an additional factor $f^c_{ijk}$ was also introduced into the above;
it takes values of $f^c_{ijk}=1$ if the collision with the impacting 
grain $R_j$ is energetic enough to disrupt the target grain $R_i$, and with
$f^c_{ijk}=0$ if the collision does not disrupt grain $R_i$. The threshold
for collisional disruption is given later in Section \ref{disruption}.

The leftmost sum in the above runs over streamlines that are composed of dust of
radii $R_j$ that have size parameters 
$\beta_j=\beta_{max}(R_{min}/R_j)$ produced
by the planetesimal ring that has radius $r_j$; their orbit elements $a_j$ and
$e_j$ are given by Equations (\ref{a}) and (\ref{e}). These dust grains have
$N_\beta$ possible sizes, and there are $N_r$ such planetesimal rings,
so $1\le j\le N_\beta N_r$. The middle sum then accounts for the
$N_l$ sites in each planetesimal ring that produce
dust having relative
longitudes $\tilde{\omega}_k = \tilde{\omega}_j - \tilde{\omega}_i$,
so $1\le k\le N_l$. Also keep in mind that pairs of orbits cross at two sites
$r_{ijk}=r_1$ and $r_{ijk}=r_2$ whose contributions are accounted for
by the rightmost sum.

But first a comment on differently-sized dust grains that are
produced at the same site within the same planetesimal ring. Such grains have
the same longitude of periapse, $\tilde{\omega}_i=\tilde{\omega}_j$,
and thus will re-encounter each other at periapse with the same velocity
$\mathbf{v}_i=\mathbf{v}_j$. This also makes $\sin\phi_{ijk}=0$, which
might appear problematic because the collision probability
$\alpha_{ijk}$ would seem to be singular
there. But keep in mind that grains $i$ and $j$ have the same velocity at
periapse, namely, the planetesimal ring's velocity. Consequently, these grains
have zero relative velocity at this particular site,
so there is no chance for collisional fragmentation, and $\alpha_{ijk}$
is set to zero in this instance.

A dust grain's velocity is
$v_j=(2\pi r_{out}/T_{out})\sqrt{(1-\beta_j)(2r_{out}/r_{ijk} - r_{out}/a_j)}$
where $r_{ijk}$ is the distance from the star where streamlines
$i$ and $j$ cross. Since
$T_i/T_{out}=(a_i/r_{out})^{3/2}/\sqrt{1-\beta_i}$, the $\alpha_{ijk}$ in
the above then becomes
\begin{eqnarray}
    \label{alpha_ijk}
    \alpha_{ijk}  = \frac{\beta_{max}^2f^c_{ijk}}{4 I\sin\phi_{ijk}}
        \sqrt{\frac{1-\beta_i}{2a_j/r_{ijk} - 1}}
        \left(\frac{\beta_i + \beta_j}{\beta_i\beta_j}\right)^2
        \left(\frac{r_{out}^2}{r_{ijk}a_j}\right)
        \left(\frac{r_{out}}{a_i}\right)^{3/2}
        \left(\frac{R_{min}}{r_{out}}\right)^{2}.
\end{eqnarray}
This quantity is the collision probability density, in the sense that
$\alpha_{ijk}n_j$ is the probability per time $T_{out}$ that a dust grain in
streamline $i$ suffers a collision with the dust in streamline $j$
at site $r_{ijk}$.

%%%%%%%%%%%%%%%%%%%%%%%%%%%%%%%%%%%
% orbit crossing
%%%%%%%%%%%%%%%%%%%%%%%%%%%%%%%%%%%
\subsubsection{orbit crossing sites}
\label{orbit crossing}

The $r_{ijk}$ in Equation (\ref{alpha_ijk}) is one of two sites where
orbit $i$ crosses orbit $j$, and it is sensitive to the orbits'
relative longitude of periapse $\tilde{\omega}_k=\tilde{\omega}_j - \tilde{\omega}_i$.
To solve for $r_{ijk}$, rotate the coordinate system so that the
$\mathbf{\hat{x}}$ axis points to orbit $i$'s periapse. Requiring
the two orbital ellipses to intersect yields
\begin{eqnarray}
    \label{r_ijk}
    r_{ijk} = \frac{p_i}{1+e_i\cos\theta} 
        = \frac{p_j}{1+e_j\cos(\theta-\tilde{\omega}_k)},
\end{eqnarray}
where $p_i=a_i(1-e_i)$ is the orbit's semilatus rectum
and $\theta$ is measured from the $\mathbf{\hat{x}}$ direction. Using
$\cos(\theta-\tilde{\omega}_k)=\cos\theta\cos\tilde{\omega}_k 
+ \sin\theta\sin\tilde{\omega}_k$ allows
Equation (\ref{r_ijk}) to be written as $A\cos\theta - B\sin\theta + C = 0$,
where $A=e_i - pe_j\cos\tilde{\omega}_k$, $B=pe_j\sin\tilde{\omega}_k$,
$C=1-p$, and $p=p_i/p_j$. Solving for $\sin\theta^2 = 1 - \cos\theta^2$
then yields $\cos\theta^2 +2D\cos\theta + E=0$
where $D=AC/(A^2+B^2)$ and $E=(C^2-B^2)/(A^2+B^2)$, which has solution
\begin{mathletters}
    \label{cos_theta}
    \begin{eqnarray}
        \cos\theta &=& -D \pm \sqrt{D^2-E}\\
        \mbox{and}\quad \sin\theta &=& (A\cos\theta + C)/B.
    \end{eqnarray}
\end{mathletters}
Solve Equations (\ref{cos_theta}) for the two $\theta$'s and insert those into
$r_{ijk}=p_i/(1+e_i\cos\theta)$ to get the radial distances
for where these orbits cross.
Then rotate the coordinate system back,
$\theta \rightarrow\theta + \tilde{\omega}_i$, to get the longitudes where
these orbits cross. Note that Equations (\ref{cos_theta}) are real
when $D^2\ge E$, which is the requirement for orbits $i$ and $j$
to cross.

To evaluate the $\sin\phi_{ijk}$ in Equation (\ref{alpha_ijk}),
the dust grain's radial velocity $v_{i,r}$ and tangential velocity $v_{i,\theta}$
need to be evaluated at the collision site. Those velocity components are
\begin{mathletters}
    \label{velocities}
    \begin{eqnarray}
        \label{v_r}
        v_{i,r} &=& \frac{e_ia_i\Omega_i}{\sqrt{1-e_i^2}}
            \sin(\theta-\tilde{\omega}_i)\\
        \label{v_theta}
        \mbox{and}\quad v_{i,\theta} &=&  
            \frac{a_i\Omega_i}{\sqrt{1-e_i^2}}\left[1 + 
            e_i\cos(\theta-\tilde{\omega}_i)\right]
    \end{eqnarray}
\end{mathletters}
where $\Omega_i=2\pi/T_i$ is the grains' mean motion.
These velocity components are then used to obtain
$\sin\phi_{ijk}=|\mathbf{v}_i\times\mathbf{v}_j|/v_iv_j
=|v_{i,r}v_{j,\theta} - v_{i,\theta}v_{j,r}|/v_iv_j$ where
$v_i$ is the grain's total velocity.

Note that the above ignored the grain's small vertical velocity
$v_{i,z}$, which is of order $Iv_i$. This is appropriate since $I\ll1$.
However the following Section will still need an estimate of the dust grains'
relative velocity in the vertical direction
when determining whether collisions are destructive.
And since the model being developed here does not faithfully follow
the dust grains' vertical motions, it is assumed here that any two pairs
of dust grains have a relative vertical speed that is equal to their rms value,
$|v_{i,z} - v_{j,z}|=\sqrt{(Iv_i)^2 + (Iv_j)^2}$.

%%%%%%%%%%%%%%%%%%%%%%%%%%%%%%%%%%%
% orbit crossing
%%%%%%%%%%%%%%%%%%%%%%%%%%%%%%%%%%%
\subsubsection{disruption threshold}
\label{disruption}

Now derive the threshold for a dust grain's collisional disruption.
The target dust grain has mass $m_i$ and velocity $\mathbf{v}_i$,
and it is struck by an impacting dust grain of mass $m_j$ and velocity $\mathbf{v}_j$.
The two dust grains' relative velocity just prior to impact is
$\mathbf{v}_r=\mathbf{v}_j - \mathbf{v}_i$, and the specific work that the impactor
must do on the target grain in order to disrupt it is $Q^\star$.
For simplicity, assume that any debris produced by the disruption of
the target grain has zero dispersion velocity, which then provides the minimum energy
that is needed for disruption. In the target dust grain's rest frame,
the system's kinetic energy is $\onehalf m_j v_r^2$ just prior to impact.
It is shown below that
most collisions will shatter both the target and the impactor, so
the post-impact debris will have a total energy
$E=\onehalf m_j v_r^2 - Q^\star(m_i + m_j)$
where $Q^\star m_i$ is the work that must do
in order to shatter grain $m_i$. But this energy can also be written
as $E=\onehalf m_i v_i'^2 + \onehalf m_j v_j'^2$ where $v_i'$ is the
post-impact speed of the debris from grain $m_i$, and $v_j'$ is the post-impact
speed of mass $m_j$ in this reference frame. Momentum conservation also requires that
$m_j v_r = m_i v_i' + m_j v_j'$. Solving the energy and momentum equations
simultaneously then provides the post-impact speed of debris $m_i$,
\begin{equation}
    \label{v_i'}
    v_i' = \frac{m_j v_r}{m_i + m_j}\left[ 1 +
        \sqrt{1 - \frac{2(m_i+m_j)^2Q^\star}{m_im_jv_r^2}} \right].
\end{equation}
Keep in mind that this is the speed of debris in the reference frame
where mass $m_i$ was stationary just before impact. Consequently,
the speed of that debris in the inertial reference frame is
\begin{equation}
    \label{v_recoil}
    \mathbf{v}_i + v_i'\mathbf{\hat{v}}_r
\end{equation}
where $\mathbf{\hat{v}}_r$
is the unit vector that points in the direction of $\mathbf{v}_r$.

The collision is energetic enough to disrupt grain $m_i$
when Equation (\ref{v_i'}) is real, which requires
$v_r^2\ge 2(m_i + m_j)^2Q^\star/m_im_j$. Assuming all grains are spheres
of similar density, this requirement becomes
\begin{equation}
    \label{disrupt_threshold}
    v_r^2>  \frac{2Q^\star(\beta_i^3 + \beta_j^3)^2}{(\beta_i\beta_j)^3}
\end{equation}
When the above is satisfied, the collision is fast enough
to destroy the target grain $\beta_i$,
which is accounted for in Equation (\ref{alpha_ijk}) by setting $f^c_{ijk}=1$,
while $f^c_{ijk}=0$ when the above is not satisfied.
In that case grain $m_i$ will recoil from the collision and enter a new orbit
about the star. However, nearly all
collisions result in the disruption of grain $m_i$, so it is safe to ignore
these rare non-disruptive collision events.

Unfortunately, the dust grains' collisional specific
energy $Q^\star$ is rather uncertain, and likely depends on the target
grain's size $R_i$ \citep{HGH02}. Another difficulty is that experimental measurements
of $Q^\star$ only extend down to cm-sized targets \citep{H94}.
However models of observed debris disks suggest $Q^\star\sim3\times10^6$
ergs/gm \citep{WSS07}, while \cite{SC06} recommend a nominal value of
$Q^\star\sim10^7$ ergs/gm if the dust grains are rocky.
But if circumstellar dust grains are icy,
then values $\sim100$ smaller are possible \citep{HGH02}.
\cite{SC06} also note that sandblasting machines accelerate destructive
particles to $v\sim100$ m/sec, which would suggest an upper limit of 
$Q^\star\sim v^2 \lesssim 10^8$ erg/gm. However, all collisions in this model
are disruptive when $Q^\star<10^6$ ergs/gm, so the following
will only consider the interval $10^6<Q^\star<10^8$ ergs/gm.
And in the example model of Section \ref{example}, $99.7\%$ of all collisions
are destructive when $Q^\star=10^7$ ergs/gm while 
$89\%$ of collisions are destructive when $Q^\star=10^8$ ergs/gm.

%%%%%%%%%%%%%%%%%%%%%%%%%%%%%%%%%%%
% Evolution
%%%%%%%%%%%%%%%%%%%%%%%%%%%%%%%%%%%
\subsection{the debris disk's time evolution}
\label{evolution}

The debris disk's dust abundance
evolves due to dust production by the planetesimal rings minus
losses due to collisions among dust grains.
To quantify this, let $p_i=p(\beta_i, r_i)$ be the rate that one
site in planetesimal ring $r_i$ produces dust of size $\beta_i$ that
gets injected into streamline $i$. The abundance $n_i$ of dust in
streamline $i$ then evolves according to the rate equation
\begin{eqnarray}
    \label{dN/dt_1}
    \frac{dn_i}{dt} = p_i - {\cal R}_i
        = p_i - \frac{n_i}{T_{out}}\sum_{j=1}^{N_{r\beta}}
        \sum_{k=1}^{N_l}\sum_{r_1}^{r_2}\alpha_{ijk}n_j,
\end{eqnarray}
where the left term accounts for dust production and the right term
accounts for collisional destruction, Equation (\ref{dN^c}). 
Index $i$ refers to the dust that reside in the target streamline $i$,
while $j$ refers to dust grains in the impacting streamline $j$
that have a relative longitude of periapse
$\tilde{\omega}_k = \tilde{\omega}_j - \tilde{\omega}_i$.
Although the model consists of $N_s=N_rN_lN_\beta$
streamlines, there are only $N_{r\beta}=N_rN_\beta$ equations in
Equation (\ref{dN/dt}) that are distinct, since the $N_l$ streamlines that are
generated in the same planetesimal ring $r_i$ with the same dust size
$\beta_i$ have an identical evolution due to the system's azimuthal symmetry.

The total number of dust grains of size $\beta_i$ that are generated by
planetesimal ring $r_i$ is $N_i=N_ln_i$, where $N_l$ is the
number of dust-producing sites in a planetesimal ring. Similarly, the total
rate at which ring $r_i$ produces dust of size $\beta_i$ is $P_i=N_lp_i$.
Multiplying Equation (\ref{dN/dt_1}) by $N_l$ and setting
\begin{eqnarray}
    \label{alpha_bar}
    \bar{\alpha}_{ij} = \frac{1}{N_l}\sum_{k=1}^{N_l}\sum_{r_1}^{r_2}\alpha_{ijk}
\end{eqnarray}
then provides the rate at which $N_i$ evolves over time,
\begin{eqnarray}
    \label{dN/dt}
    \frac{dN_i}{dt} = P_i - \frac{N_i}{T_{out}}\sum_{j=1}^{N_{r\beta}}
        \bar{\alpha}_{ij}N_j.
\end{eqnarray}
Here, $\bar{\alpha}_{ij}$ is the mean probability per time $T_{out}$
that a grain of size $\beta_i$ and orbit elements $a_i, e_i$ collides with 
a grain of size and orbit $\beta_j, a_j, e_j$. Note that this is a mean
probability since Equation (\ref{alpha_bar}) averages probabilities
over all possible longitudes of periapse
$\tilde{\omega}_k = \tilde{\omega}_j - \tilde{\omega}_i$.

This model is also going to assume that, when dust grains collide, the resulting
dust fragments are so small that they are unbound and driven away by radiation
pressure. This assumption simplifies the problem enormously since, if it 
where not true, then the number of streamlines in the debris disk
would grow exponentially as collisions beget additional dusty streamlines that
collide with other dust and generate even more streamlines. The validity of
this assumption is confirmed later in Section \ref{fragmentation}.

%%%%%%%%%%%%%%%%%%%%%%%%%%%%%%%%%%%
% Two sizes
%%%%%%%%%%%%%%%%%%%%%%%%%%%%%%%%%%%
\subsubsection{two-component model}
\label{two_sizes}

Now consider a rather simple system, that of a
single planetesimal ring that manufactures
dust having only two sizes, small (S) grains having radii near the blow-out radius
$R_{min}$, and large (L) grains of radii $R_L\gg R_{min}$. In the resulting debris disk,
nearly all collisions are with the small grains, due to their much greater abundance
in the disk, $N_S\gg N_L$, and their much greater production rates $P_S\gg P_L$.
Because small grains are only colliding with
other small grains, their abundance varies as
\begin{eqnarray}
    \label{dN/dt_singlesize}
    \frac{dN_S}{dt} = P_S - \frac{\bar{\alpha}_{SS}N_S^2}{T_{out}}
\end{eqnarray}
according to Equation (\ref{dN/dt}),
where $\bar{\alpha}_{SS}$ is the probability density
for collisions among small grains. The solution is
\begin{eqnarray}
    \label{tanh}
    N_S(t) = N_S^{eq}\tanh(t/t_S),
\end{eqnarray}
where 
\begin{eqnarray}
    \label{N_S^eq}
    N_S^{eq}=\sqrt{\frac{P_ST_{out}}{\bar{\alpha}_{SS}}}
        \qquad\mbox{and}\qquad
        t_S = \sqrt{\frac{T_{out}}{\bar{\alpha}_{SS}P_S}}=\frac{N_S^{eq}}{P_S}.
\end{eqnarray}
Here, $N_S^{eq}$ is the total number of small dust
grains that result when the disk settles into equilibrium, 
which occurs at time $t\gg t_{S}$ when $\tanh(t/t_{S})\rightarrow1$
and $dN_S/dt\rightarrow0$.
Note that $t_{S}$ is also the small dust grains' collisional lifetime.
Evidently the small grains' equilibrium 
abundance as well as their optical depth will vary as $N_S^{eq}\propto P_S^{1/2}$
where $P_S$ is their production rate, while the timescale for the disk to
settle into collisional equilibrium varies as $t_S\propto P_S^{-1/2}$.

The large grains' abundance $N_L$ also evolves according to Equation (\ref{dN/dt}),
which becomes $dN_L/dt = P_L - \bar{\alpha}_{LS}N_LN_S/T_{out}$ when collisions
with small grains dominate. And when the disk has settled into collisional equilibrium,
$dN_L/dt = 0$, so the equilibrium abundance of large grains is
\begin{eqnarray}
    \label{N_L^eq}
    N_L^{eq}=\frac{P_LT_{out}}{\bar{\alpha}_{LS}N_S^{eq}}
\end{eqnarray}
where $\bar{\alpha}_{LS}$ is the large grain's collisional probability density.

Although this two-component treatment might seem too simple for quantitative work,
it is illustrative, because
a real disk's optical depth and collision rates are often dominated by
the disk's smallest grains that have size parameters close to $\beta_{max}$.
In that case, the disk's collisional equilibrium timescale 
and its optical depth would all vary
with the square root of the planetesimal's dust production rate. Equations
(\ref{N_S^eq}--\ref{N_L^eq}) 
are also useful because they provide a convenient test
of the more general debris disk model that is
developed below in Section \ref{size_distribution}; using that more general
model to simulate a two-component debris yields results that agree
with Equations (\ref{tanh}--\ref{N_L^eq}) to within $0.04\%$.

%%%%%%%%%%%%%%%%%%%%%%%%%%%%%%%%%%%
% scale-invariant evolution
%%%%%%%%%%%%%%%%%%%%%%%%%%%%%%%%%%%
\subsubsection{scale-invariant evolution}
\label{size_distribution}

Now consider a more realistic scenario where the system's
planetesimal rings produce dust whose rates are power-laws in the dust
size parameter $\beta_i$ and planetesimal ring radius $r_i$ such that
$P_i =P_0(\beta_i/\beta_{max})^q(r_i/r_{out})^c$.
It will also be convenient to first convert Equation
(\ref{dN/dt}) into a dimensionless system of equations via the substitutions
\begin{eqnarray}
    \label{dimensionless}
    t = T_0 t^\star  \qquad\mbox{and}\qquad
        N_i(t) = N_0\left(\frac{\beta_i}{\beta_{max}}\right)^q 
            \left(\frac{r_i}{r_{out}}\right)^c N_i^\star(t^\star)
\end{eqnarray}
where $t^\star=t/T_0$ is  dimensionless time coordinate
and $N_i^\star(t^\star)=N_i(t)P_0/P_iN_0$ is the scaled abundance
of grains of size $\beta_i$ produced by ring $r_i$.
The collision probability density will
also be written as
$\bar{\alpha}_{ij} = (\bar{\alpha}^\star_{ij}/I)(\beta_j/\beta_{max})^{-q}
(r_j/r_{out})^{-c}(R_{min}/r_{out})^2$
where the scaled collision probability is
\begin{eqnarray}
    \label{alpha_star}
    \bar{\alpha}^\star_{ij} = \frac{1}{N_l}\sum_{k=1}^{N_l}\sum_{r_1}^{r_2}
         \frac{\beta_{max}^2f^c_{ijk}}{4\sin\phi_{ijk}}
            \sqrt{\frac{1-\beta_i}{2a_j/r_{ijk} - 1}}
            \left(\frac{\beta_i + \beta_j}{\beta_i\beta_j}\right)^2
            \left(\frac{\beta_j}{\beta_{max}}\right)^{q}
            \left(\frac{r_{out}^2}{r_{ijk}a_j}\right)
            \left(\frac{r_{out}}{a_i}\right)^{3/2}
            \left(\frac{r_j}{r_{out}}\right)^{c}.\hspace*{3ex}
\end{eqnarray}
In the above, the constant timescale $T_0$ and population scale-factor $N_0$ are 
\begin{mathletters}
    \label{T0n0}
    \begin{eqnarray}
        \label{T0}
        T_0 &=& \sqrt{\frac{IT_{out}}{P_0}}\left(\frac{r_{out}}{R_{min}}\right)\\
        \label{n0}
        \mbox{and}\quad N_0 &=& P_0T_0 = 
            \sqrt{IP_0T_{out}}\left(\frac{r_{out}}{R_{min}}\right).
    \end{eqnarray}
\end{mathletters}
Inserting this into Equation (\ref{dN/dt}) then yields a system of equations
\begin{eqnarray}
    \label{dn/dt}
    \frac{dN^\star_i}{dt^\star} = 1 - 
        N^\star_i\sum_{j=1}^{N_{r\beta}}\bar{\alpha}^\star_{ij}N^\star_j
\end{eqnarray}
that is not just dimensionless, but also scale invariant. This is very handy,
since Equation (\ref{dn/dt}) need only be solved once in order to apply the
result to a variety of systems that might have a range of planetesimal 
radii $r_{out}$ orbiting stars of varied masses $M_\star$ and luminosities $L_\star$.
Note, however, that the rescaled collision
probability  $\bar{\alpha}^\star_{ij}$ does depend of the dust power laws
$q$ and $c$, the dust grains' specific energy for collisional disruption $Q^\star$,
and the planetesimal disk's radial width $r_{in}/r_{out}$
(Equation \ref{alpha_star}), so any changes to those parameters does require
solving Equation (\ref{dn/dt}) again.

\begin{figure}[h]
    \epsscale{0.68}
    \vspace*{-1ex}\plotone{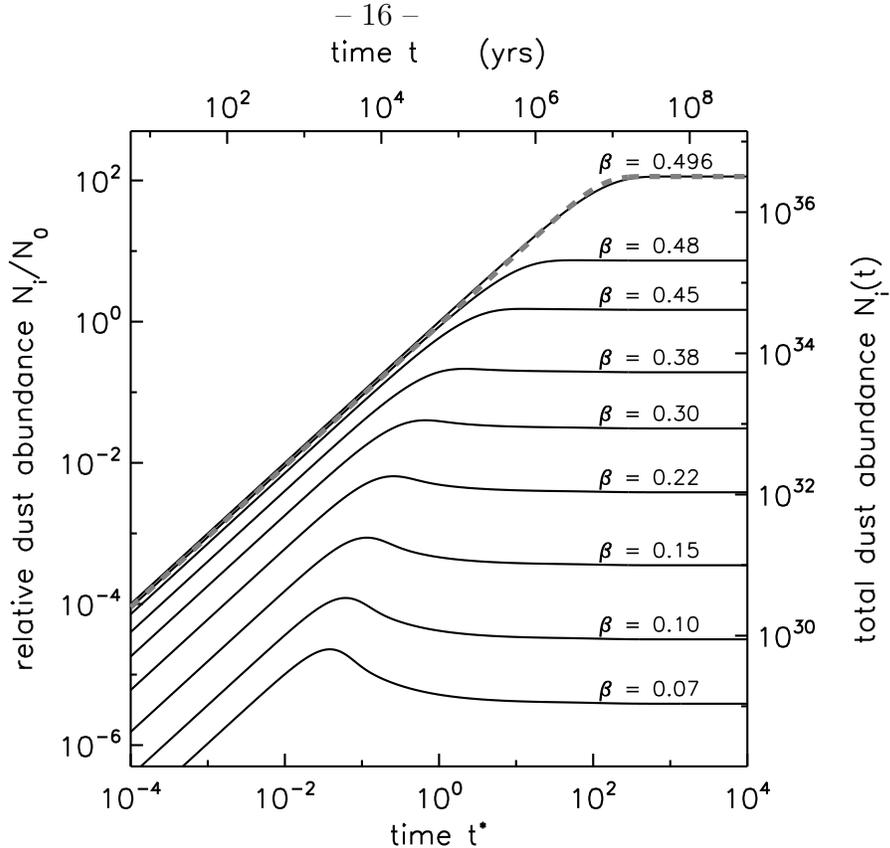}\vspace*{-4ex}
    \figcaption{
        \label{n(tau)_figure}
        Equation (\ref{dn/dt}) is solved for the scaled dust abundances
        $N^\star_i(t^\star)$ assuming that the dust is produced by a narrow planetesimal
        ring ($N_r=1$) with a $q=3.5$ size distribution. The scaled dust abundances 
        are then converted into absolute dust abundances $N_i(t^\star)/N_0$
        and plotted versus dimensionless time $t^\star$ (left and lower axes).
        This model also assumed that the dust are weak,
        $Q^\star=10^6$ ergs/gm, so all collisions
        disrupt the dust. Abundances for
        nine selected streamlines having the indicated size parameter $\beta$
        are shown above; not shown are the model's 191 other curves that
        behave similarly. The right and upper axes are scaled for an
        $r_{out}=50$ AU planetesimal ring
        in orbit about a solar-type star that produces
        dust at the rate $\dot{M}_d=10^{13}$ gm/sec with
        $\rho=1$ gm/cm$^3$, $Q_{rp}=0.96$, and $I=0.1$ radians, for which 
        $N_0 =2.8\times10^{34}$ grains and $T_0=5.6\times10^4$ yrs.
        The dashed grey curve is Equation (\ref{tanh}) with an equilibrium timescale
        $T_{eq}=7.1$ Myrs.                                                       
    }
\end{figure}

Equation (\ref{dn/dt}) is a coupled set of nonlinear differential equations
whose initial conditions are $N^\star_i(0)=0$. These equations
are easily solved numerically for each streamline's relative
abundance $N^\star_i(t^\star)$ using a Runga-Kutta algorithm.
That solution to Equation (\ref{dn/dt})
then provides the dust grains' relative abundances
$N_i(t)/N_0=(P_i/P_0)N^\star_i(t^\star)$.
A sample calculation is shown in Figure \ref{n(tau)_figure},
which plots the relative abundances 
$N_i/N_0$ versus dimensionless time $t^\star$
for dust that is generated by a narrow
planetesimal ring ($N_r=1$ and $r_{in}=r_{out}$) whose dust production has the
$q=3.5$ size distribution that expected for debris that is
 in collisional equilibrium \citep{D69}.  In this and all calculations that follow,
the simulated debris disk is composed of grains having
$N_\beta=200$ distinct sizes, with the
dust size parameters distributed over the interval $0.0652\le\beta_i\le0.497$ such
that the dust grains' radii $R_i=(\beta_{max}/\beta_i)R_{min}$
are uniformly sampled over the interval $R_{min}\le R\le R_{max}$.
Dust of each size $\beta_i$ are launched from $N_l=100$ evenly-spaced longitudes
about the planetesimal disk, so the resulting
debris disk is comprised of $N_s=N_rN_lN_\beta=2\times10^4N_r$
distinct streamlines.
The planetesimal disk that is the source of this dust is usually composed of
$N_r=3$ to 5 planetesimal rings, except $N_r=1$ is used when the dust source
is a narrow planetesimal ring, as is the case for Figure \ref{n(tau)_figure}.
The $\beta$ distribution used here results in
streamlines that range as far as  $r=150r_{out}$ from the central star,
where $r_{out}$ is the planetesimal disk's outer radius. Execution times
on a desktop PC range from 4 minutes for simulations having $N_r=1$ to
2 hours when $N_r=5$.

%Fig: n(t) figure
\newpage

Although the calculation seen in Figure \ref{n(tau)_figure} covers a fairly broad
range of $\beta$ parameters, the resulting debris disk
is nonetheless populated by dust having a fairly narrow range of sizes,
since it is only the smallish dust grains that are lofted well outside of the
birth ring by radiation pressure.
In  Figure \ref{n(tau)_figure}, the central star is solar and the dust grains
have a density $\rho=1$ gm/cm$^3$ and radiation pressure efficiency
$Q_{rp}=0.96$, so the radius of the smallest bound dust grain is
$R_{min}=1.1$ $\mu$m, and the range of dust radii that populate
the simulated debris disk is only $1.1$ $\mu$m$\le R \le 8.5\ \mu$m.
Of course, a real disk will also manufacture larger dust,
but those larger grains would be confined to the disk's innermost
region $r_{in}\le r\le 1.15r_{out}$,
so they do not alter the disk's large-scale structure at $r\gg r_{out}$.
Also, these larger grains contribute little to the 
disk's optical depth due to their slower dust production rates
and their very short collisional lifetime, so their absence from these simulations
is justified.

The parameter $P_0$ that appears in the constants $N_0$ and $T_0$
(Equations \ref{T0n0})
is related to the planetesimal disk's total dust mass production rate, which is
$\dot{M}_d=\sum_iP_i\frac{4\pi}{3}\rho R_i^3$ where the sum proceeds
over all streamlines that make up the dust disk.
Since $P_i=P_0(\beta_i/\beta_{max})^q(r_i/r_{out})^c$, this provides
$P_0=\dot{M}_d/S_{qc}m_{min}$, where $m_{min}=\frac{4\pi}{3}\rho R_{min}^3$
is the mass of the smallest bound dust grain, and the factor
$S_{qc}=\sum_i(\beta_i/\beta_{max})^{q-3}(r_i/r_{out})^c$.
Inserting this and Equation
(\ref{R_min}) into (\ref{T0n0}) then provides the scale factors
{\small
\begin{mathletters}
    \label{N0T0}
    \begin{eqnarray}
        \label{N03}
        N_0 &=& \frac{2.8\times10^{34}}{Q_{rp}^{5/2}}
            \left(\frac{\dot{M}_d}{10^{13}\mbox{ gm/sec}}\right)^{1/2}
            \left(\frac{I}{0.1\mbox{ rad}}\right)^{1/2}
            \left(\frac{L_\star}{L_\odot}\right)^{-5/2}
            \left(\frac{M_\star}{M_\odot}\right)^{9/4}\\
            &&\times
            \left(\frac{\rho}{1\mbox{ gm/cm}^3}\right)^{2}
            \left(\frac{r_{out}}{50\mbox{ AU}}\right)^{7/4}
            \left(\frac{S_{qc}}{107}\right)^{-1/2}\\
        \label{T03}
        \mbox{and}\quad T_0 &=& 5.6\times10^4Q_{rp}^{1/2}
            \left(\frac{\dot{M}_d}{10^{13}\mbox{ gm/sec}}\right)^{-1/2}
            \left(\frac{I}{0.1\mbox{ rad}}\right)^{1/2}
            \left(\frac{L_\star}{L_\odot}\right)^{1/2}
            \left(\frac{M_\star}{M_\odot}\right)^{-3/4}\\
            &&\times
            \left(\frac{r_{out}}{50\mbox{ AU}}\right)^{7/4}
            \left(\frac{S_{qc}}{107}\right)^{1/2}\mbox{ yrs}
    \end{eqnarray}
\end{mathletters}
}
where the factor $S_{qc}=107$, 320, and 530 for
the simulations having $N_r=1$, 3, and 5 
planetesimal rings with $c=0$ and $q=3.5$.
The above quantities allow one to easily 
rescale all figures shown here
for systems having alternate dust production rate $\dot{M}_d$
or the planetesimal disk's outer radius $r_{out}$, etc.
The following subsection will also show that
the time $T_{eq}$ for the resulting dust-disk to settle into an equilibrium
where dust production balances mass-loss due to collisions
is $T_{eq}\simeq130T_0$.

%%%%%%%%%%%%%%%%%%%%%%%%%%%%%%%%%%%
% ar=50 AU example
%%%%%%%%%%%%%%%%%%%%%%%%%%%%%%%%%%%
\subsubsection{example: $a_r=50$ AU birth ring}
\label{example}

To illustrate, consider a narrow ($N_r=1$) planetesimal ring
of radius $r_{out}=50$ AU in orbit
about a solar-type star. This ring will have a dust mass
production rate of $\dot{M}_d=10^{13}$ gm/sec, with dust having
a density $\rho=1$ gm/cm$^3$ and inclinations $I=0.1$ radians.
The grains are dark, with $Q_s=0.1$,
and are asymmetric light scatters having $g=0.4$, so the 
radiation pressure efficiency is $Q_{rp}=0.96$,
$N_0 = 2.8\times10^{34}$ and $T_0=5.6\times10^4$ yrs. Multiplying
the lower and left axes in Figure \ref{n(tau)_figure} by these constants
then provides the dust abundances $N_i(t)$ versus physical time $t$,
which can also be read off the right and upper axes of Figure \ref{n(tau)_figure}.
That figure shows that the smallest dust grains
having $\beta\simeq\beta_{max}$ need the most time to settle into an
equilibrium where dust production balances destruction due to collisions.
The grey dashed curve in Figure \ref{n(tau)_figure} also shows that the
abundance of the smallest dust does resembles Equation (\ref{tanh})
when the equilibrium timescale $T_{eq}=7.1\times10^6$ yrs
is chosen as the moment when the smallest
dust have reached two-thirds their equilibrium abundance.
This demonstrates that the two-component dust size model of Section \ref{two_sizes}
is in fact relevant to this kind of debris disk. And because $T_0$
and $T_{eq}$ have the same dependance on the model parameters, comparison
to Equation (\ref{T03}) shows that $T_{eq}\simeq130T_0$.

%Fig: N(t) figure
\begin{figure}[t]
\epsscale{0.75}
\vspace*{-7ex}\plotone{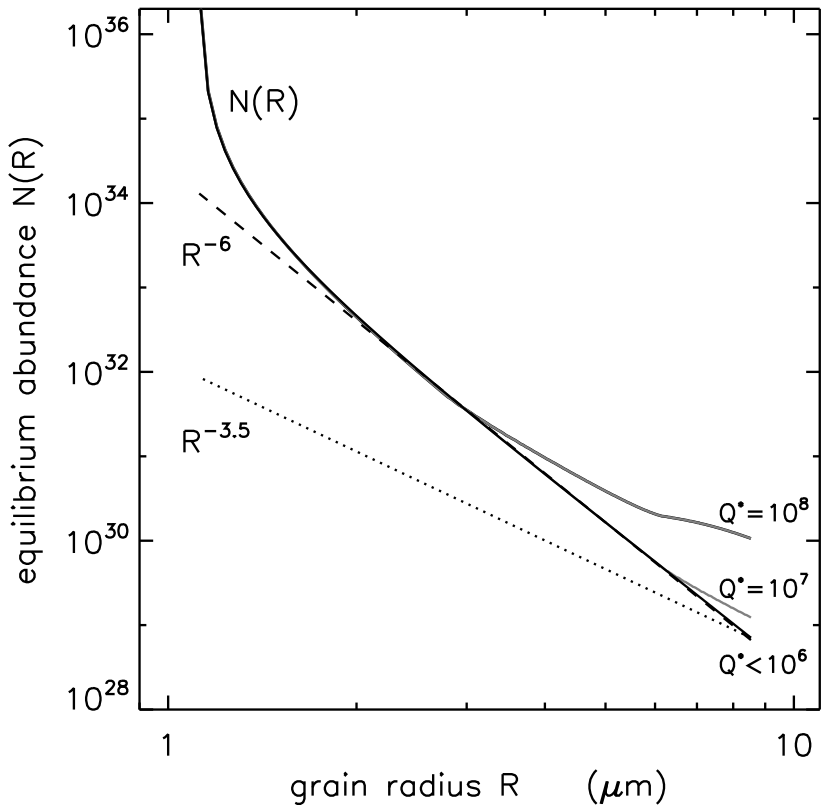}\vspace*{-4ex}
\figcaption{
    \label{N_final_figure}
    The black $N(R)$ curve is the 
    differential size distribution for the dust in the debris-disk
    model of Figure \ref{n(tau)_figure} after the disk has achieved equilibrium
    at times $t\gg T_{eq}$ assuming the dust grains are weak with
    $Q^\star<10^6$ ergs/gm.
    Grey curves are for models having stronger dust grains,
    $Q^\star=10^7$ and $10^8$ ergs/gm.
    The dashes indicates an $R^{-6}$ size distribution,
    and the dotted curve is proportional
    to the planetesimal ring's differential dust
    production rate $P(R)\propto R^{-3.5}$.
}
\end{figure}

\begin{figure}[t]
\epsscale{0.75}
\vspace*{-7ex}\plotone{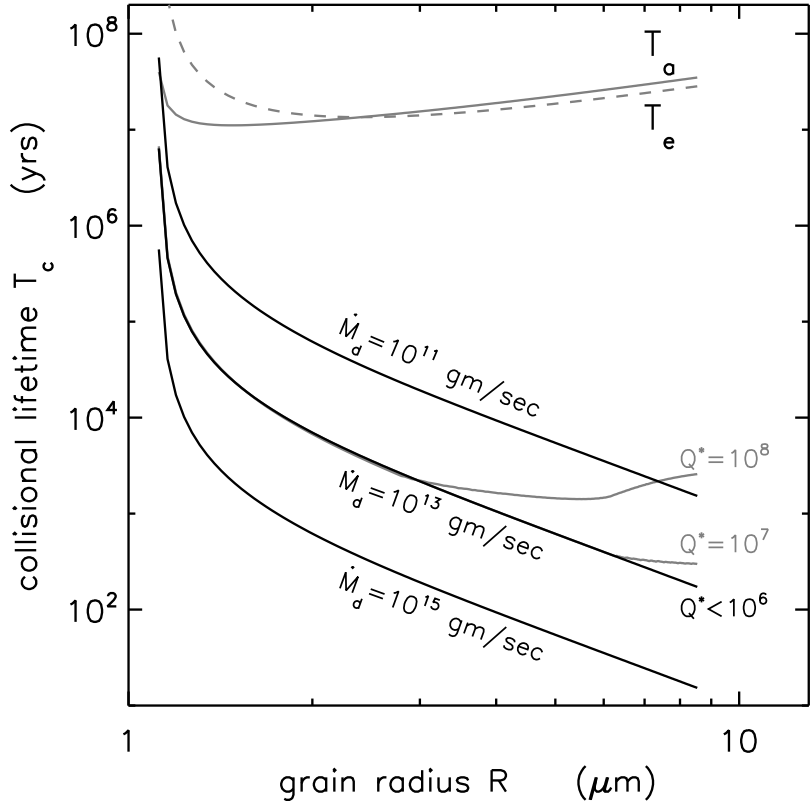}\vspace*{-4ex}
\figcaption{
    \label{T_col_figure}
    The lower middle black curve is the dust grain's collisional lifetime $T_{c}(R)$
    plotted versus dust radius $R$ for the debris-disk model of
    Figure \ref{n(tau)_figure}
    that has a dust production rate $\dot{M}_d=10^{13}$ gm/sec and strength
    $Q^\star<10^6$ ergs/gm, while the
    other black curves are for systems that are otherwise identical but with the
    indicated dust production rates $\dot{M}_d$. The grey curves are for
    stronger dust grains having $Q^\star=10^7$ and $10^8$ ergs/gm and
    $\dot{M}_d=10^{13}$ gm/sec; compare also to
    Figure \ref{N_final_figure}. The upper solid grey curve $T_a$ is the 
    timescale over which the dust grains' semimajor decays due to PR drag,
    with $T_e$ the timescale for eccentricity damping due to PR drag; see
    Equations (\ref{T_PR}).
}
\end{figure}

After a time $t\gg T_{eq}$, the debris disk will have
settled into equilibrium. The solid black curve in
Figure \ref{N_final_figure}
shows the disk's equilibrium grain size distribution $N(R)$,
which is extracted from the rightmost part of Figure \ref{n(tau)_figure}
and plotted versus grain radii $R$. That model assumes that the dust grains are
weak, $Q^\star<10^6$ ergs/gm, which means that all collisions are destructive.
Note that the debris disk's dust size
distribution $N(R)$ is significantly steeper than the $q=3.5$ size distribution
that governs the ring's dust production (the dotted curve in 
Figure \ref{N_final_figure}), with $N(R)\propto R^{-6}$ except near 
$R\simeq R_{min}$, where it has an even steeper dependance.
A peak at $R\simeq R_{min}$ is also seen in the debris disk
model\footnote{Note though that the
equilibrium dust size distribution in the \cite{KLS06} model is
`wavy'. Waves in a size-distribution occur when some process
tends to favor the rapid removal of the smallest bodies \citep{DD97},
and Figure 5 of \cite{KLS06} would suggest that next peak in our
size distribution might occur at sizes $R\sim1$ mm. However these large
grains would still have a very short collisional lifetime
(see Figure \ref{T_col_figure}), which makes their influence in the dust-disk
quite negligible.} of \cite{KLS06}. Figure \ref{N_final_figure} 
illustrates the main consequence of dust-dust collisions,
which tends to destroy the disk's larger dust grains at a faster pace.
However, when the dust grains are stronger, with $10^7<Q^\star<10^8$ ergs/gm,
then the larger grains are more resistant to collisional destruction, and they
become more abundance (Figure \ref{N_final_figure}, grey curves).

When the system is in equilibrium, the rate at which the planetesimal ring
injects dust of size $\beta_i$ into the debris disk, $P_i$, balances the rate
at which collisions remove dust from the disk ${\cal R}_i$,
so ${\cal R}_i = P_i$. Thus the collisional
lifetime of the grains in streamline $i$ is
$T_c(R_i)=N_i/{\cal R}_i = N_i/P_i$, which is plotted versus
grain radius $R$ in Figure \ref{T_col_figure}
(see the black curves
that are labeled by their dust production rates $\dot{M}_d$).
That lifetime is simply the time when the dust abundances $N_i(t)$
seen in Figure \ref{n(tau)_figure} flatten out.
As expected, the smallest grains that have size parameters $\beta$
that are just shy of $\beta_{max}=\onehalf$ are very long lived.
This is due to their orbits having very large apoapses
$Q_{apo}=r_{out}/(1-2\beta)\gg r_{out}$,
so small dust spend most of their time far from the planetesimal ring,
in regions where the disk's optical depth is low and collisions are rare.
Note also that $T_c\propto(N_0/P_0)\propto T_0$, so the
dust grains' collision timescale obeys the same scaling as Equation
(\ref{T03}), in particular with $T_c\propto \dot{M}_d^{-1/2}$.
Figure \ref{T_col_figure} also shows that nearly all dust grains
in such a debris disk have lifetimes $T_c$ that are short compared to the
age of the host star, which is typically $\sim10^7$ to $10^8$ years
\citep{MBW07}. 

When the grains are weak, $Q^\star<10^6$ ergs/gm, then the lifetime
of the larger $R\gtrsim2R_{min}$ varies at $T_c(R)\propto R^{-2.4}$
(Figure \ref{T_col_figure}). That figure also shows that
increasing the dust grains' strength $Q^\star$
increases the larger grains' longevity, without affecting the small
grains' collisional lifetime. This is because all collisions
are energetic enough to destroy all small grains.

%Fig: N(t) figure

%Fig: tau(r) at various times
\begin{figure}[t]
\epsscale{0.75}
\vspace*{-7ex}\plotone{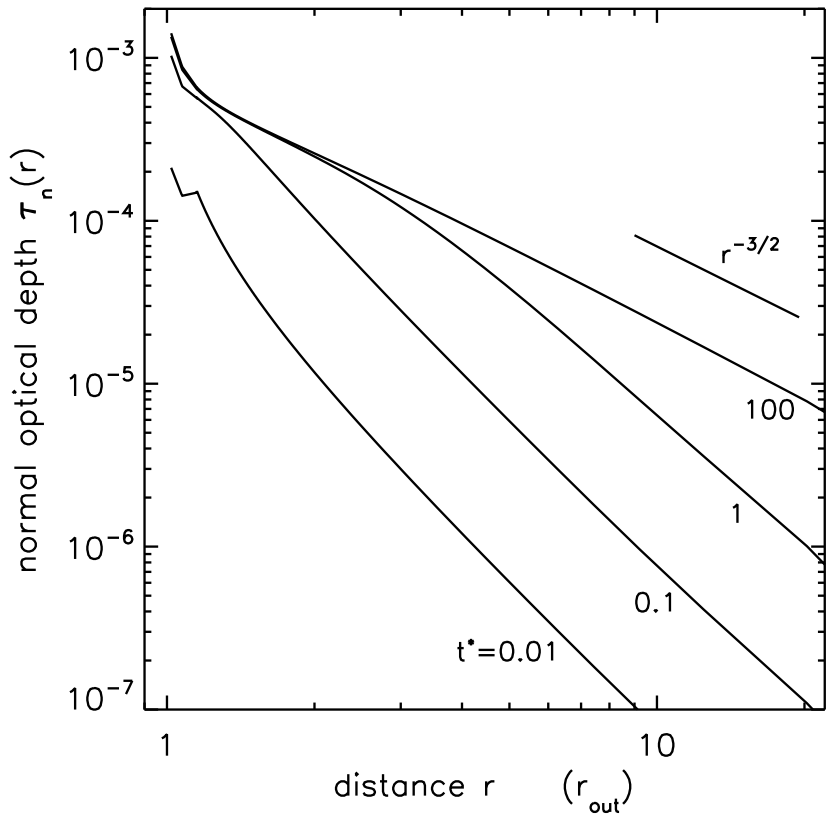}\vspace*{-4ex}
\figcaption{
    \label{tau_time_figure}
    The dust optical depth $\tau_n(r)$ is plotted versus distance $r$
    in units of the planetesimal disk's outer radius $r_{out}$
    for the model of Figure \ref{n(tau)_figure}. Shown is $\tau_n(r)$
    at dimensionless times $t^\star=0.01$, 0.1, 1, 100, and an $r^{-3/2}$ curve.
}
\end{figure}

%%%%%%%%%%%%%%%%%%%%%%%%%%%%%%%%%%%
% optical depth
%%%%%%%%%%%%%%%%%%%%%%%%%%%%%%%%%%%
\subsubsection{optical depth}
\label{depth}

To determine the model debris disk's normal optical depth $\tau_n(r)$,
it is convenient to first calculate the total dust cross section $A(r)$
that resides interior to distance $r$ from the star. That quantity is
obtained by first counting the number of dust grains $\Delta n_j(r)$ in streamline
$j$ that also lie interior $r$, which is
$\Delta n_j(r)=\int\lambda_j d\ell$ where the integration runs
along the stretch of streamline interior to $r$. This becomes
a trivial integral over time $t$ after noting
$\lambda_j d\ell=\lambda_j v_jdt$ where $\lambda_j=n_j/v_jT_j$ is the dust
grain's linear density, and $n_j$ is the total number
of grains in streamline $j$ that have velocity $v_j$ and orbit period $T_j$.
Consequently, $\Delta n_j(r)=2n_jt_j(r)/T_j$
where $t_j(r)$ is the time for dust in streamline $j$ to travel from periapse
to distance $r$. And since $n_i=N_i/N_l$ where $N_i$ is the total number of
dust grains in the disk that have the same sizes and orbits,
the total cross section for grains of radii $R_i=(\beta_{max}/\beta_i)R_{min}$
that reside interior to $r$ is $A_i(r) = N_l\Delta n_i(r)\pi R_i^2 = 
2N_i(t_i/T_i)(\beta_{max}/\beta_i)^2\pi R_{min}^2$.
The total dust cross section interior to $r$ is then $A(r)=\sum_i A_i(r)$
where the sum proceeds over all streamlines whose dust have distinct sizes
$\beta_i$ produced by planetesimal ring $r_i$.
The ratio $t_i/T_i$ in the above is obtained by solving
$r=a_i(1-e_i\cos E)$ for the dust grains' eccentric anomaly $E(r)$,
which is then inserted into Kepler's equation to
obtain $t_i/T_i = (E-e_i\sin E)/2\pi$. Also note that the differential
$\Delta A =(\partial A/\partial r)\Delta r$
is the dust cross section that resides in an annulus of radius $r$
and width $\Delta r$. Since the disk's normal optical depth $\tau_n$
is simply the surface density of dust cross section, that quantity is
\begin{eqnarray}
    \label{tau}
    \tau_n(r) &=& \frac{\Delta A}{2\pi r\Delta r} =
        \frac{1}{2\pi r}\frac{\partial A}{\partial r},
\end{eqnarray}
which is easily calculated by differentiating $A(r)$ numerically.

Figure \ref{tau_time_figure} plots the dust optical depth $\tau_n(r)$ versus distance
$r$ from the star at dimensionless times $t^\star=0.01$, 0.1, 1, and 100
for the model of Figure \ref{n(tau)_figure}.
As the figure shows, it is the outer portion of the disk that is
populated at later times by dust. This is due to the outer disk being composed of
smaller dust that, according to Figure \ref{n(tau)_figure}, 
are the last to arrive at a collisional balance.
Also note that when the disk has settled into equilibrium at
times $t^\star\gg1$, the outer disk at $r\gg r_{out}$ has an optical depth
$\tau_n(r)\propto r^{-3/2}$, which is in agreement with what \cite{SC06}
call a type B debris disk.

%Fig: tau(r) figure
\begin{figure}[t]
\epsscale{0.75}
\vspace*{-7ex}\plotone{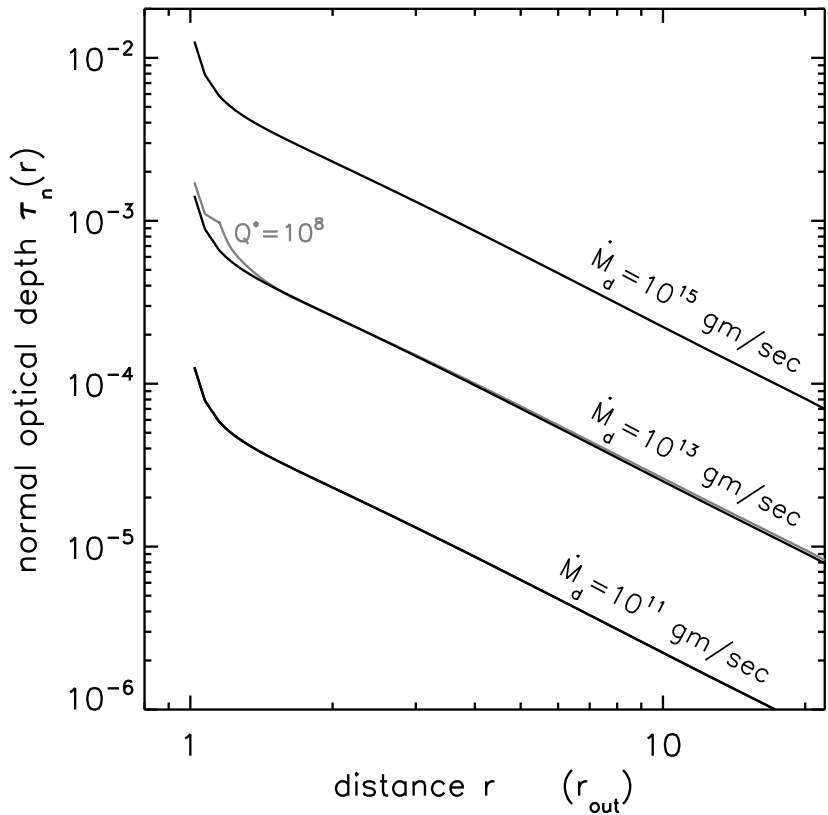}\vspace*{-4ex}
\figcaption{
    \label{tau_figure}
    Black curves give the debris disk's equilibrium optical depth $\tau_n(r)$
    versus radial distance $r$
    for dust generated in a narrow debris disk of radius $a_r$. 
    Model parameters are identical
    to those adopted in Figure \ref{n(tau)_figure} except that different
    dust production rates are considered, $\dot{M}_d=10^{11}$, $10^{13}$,
    and $10^{15}$ gm/sec, and weak grains having $Q^\star<10^6$ ergs/gm,
    while the one grey curve assumes strong dust having  $Q^\star=10^8$ ergs/gm.
}
\end{figure}

The black curves in Figure \ref{tau_figure}
show the equilibrium optical depth $\tau_n(r)$ for simulated debris
disks that are identical to the one considered in Figure \ref{n(tau)_figure},
except that different dust production rates $\dot{M}_d$ are considered.
These curves assume the dust grains are weak, with $Q^\star<10^6$ ergs/gm.
As expected, these optical depths vary as $\dot{M}_d^{1/2}$ (see Equation
\ref{N03}). The one grey curve there shows the optical depth of a disk
composed of strong dust having $Q^\star=10^8$ ergs/gm. That disk shows
a slight overdensity near the planetesimal ring, and is due to an
excess of larger grains in low-eccentricity orbits (see grey curves in
Figures \ref{N_final_figure} and \ref{T_col_figure}).

\begin{figure}[t]
\epsscale{0.75}
\vspace*{-7ex}\plotone{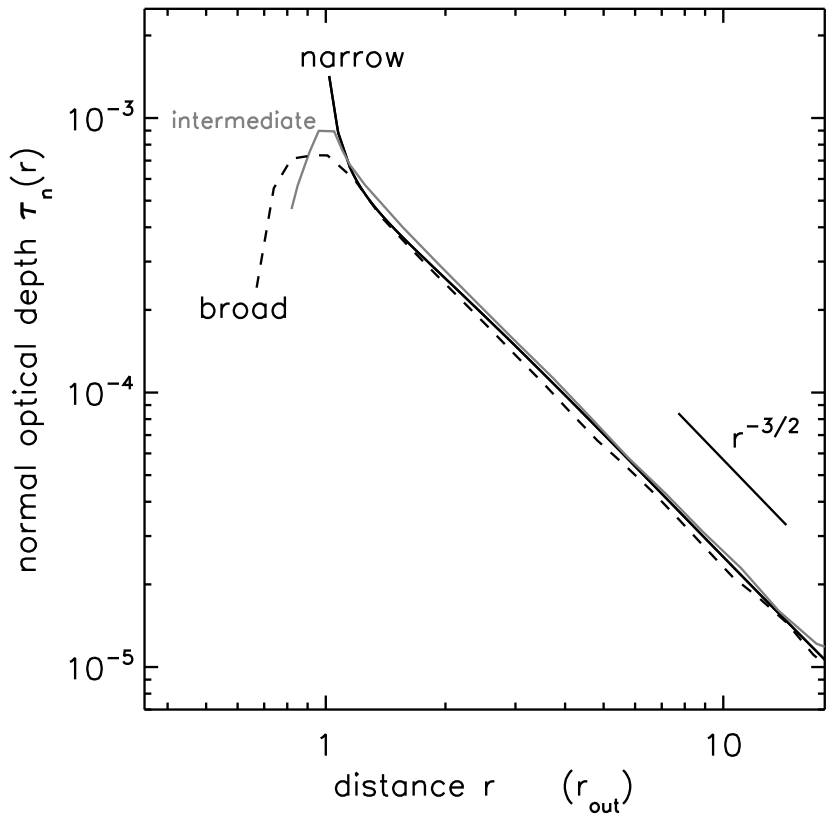}\vspace*{-4ex}
\figcaption{
    \label{tau_disk_figure}
    The normal optical depth $\tau_n(r)$ for three debris disks.
    One is generated by a narrow
    planetesimal ring of radius $r_{out}$ (black curve), 
    another by an intermediate-width planetesimal ring
    whose inner radius $r_{in}=0.8r_{out}$ (grey curve), and the third
    due to a broad planetesimal disk having $r_{in}=0.5r_{out}$ (dashed curve).
    The narrow planetesimal disk is represented by one planetesimal ring ($N_r=1$),
    while the intermediate and broad disks use $N_r=3$ and $N_r=5$
    planetesimal rings, respectively. All simulations have
    $c=0$, so the dust production rate is independent
    of distance $r$ in the planetesimal disk.
    Model parameters are otherwise identical
    to those adopted in Figure \ref{n(tau)_figure}. Also shown is an $r^{-3/2}$ curve.
}
\end{figure}

The sharp peak in $\tau_n(r)$ at $r\simeq r_{out}$ seen in
Figure \ref{tau_figure} is due to
the narrow width of the planetesimal ring that
is the source of this dust. However, Figure \ref{tau_disk_figure} shows
that peak broadens when the planetesimal ring's
radial width is increased. 

%Fig: tau(r) for ring & disk figure

%%%%%%%%%%%%%%%%%%%%%%%%%%%%%%%%%%%
% surface brightness 
%%%%%%%%%%%%%%%%%%%%%%%%%%%%%%%%%%%
\subsubsection{disk surface brightness}
\label{brightness}

This subsection will calculate the surface brightness of starlight
that a simulated debris disk will scatter towards an observer that
views the disk at optical wavelengths.
In the following, flux $F$ is the power per area in the incident or scattered
radiation, intensity $I$ is the power per solid angle in a pencil-beam of radiation,
and the surface brightness $B$ is the radiation
beam's power per area per solid angle.

%Fig: LOS figure
\begin{figure}[t]
\epsscale{0.75}
\vspace*{-8ex}\plotone{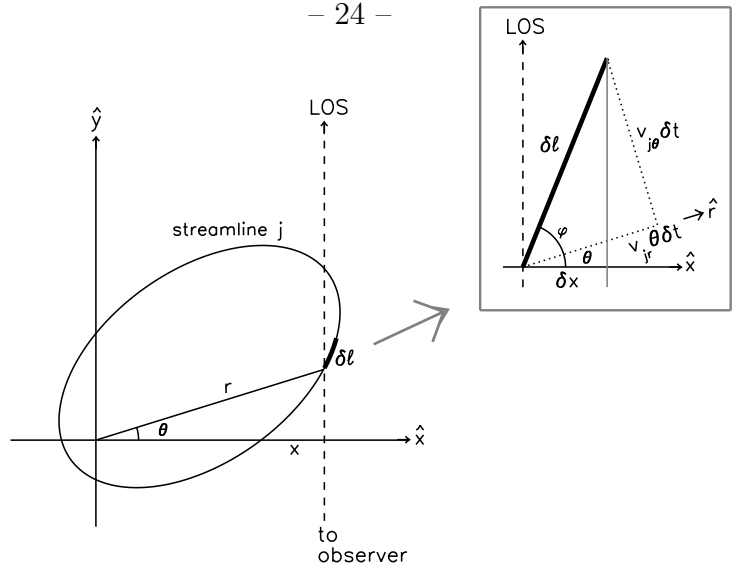}\vspace*{-3ex}
\figcaption{
    \label{los_figure}
    The observer's line-of-sight (LOS) passes through streamline $j$ and
    intersects segment $\delta\ell$ at polar coordinates $(r, \theta)$;
    that segment also lies at a projected
    distance $x$ from the central star at the origin.
    The insert zooms in on segment $\delta\ell$, which has a projected
    length $\delta x$ when viewed by the observer. The dust in that segment
    will have traveled a distance $\delta\ell=v_j\delta t$ in time  $\delta t$,
    which corresponds to radial and tangential displacements $v_{j,r}\delta t$
    and $v_{j,\theta}\delta t$, so the angle $\varphi$ satisfies
    $\sin\varphi=v_{j,\theta}/v_j$ and $\cos\varphi=v_{j,r}/v_j$. Also note
    that the scattering angle $\phi$ is the angle between the radial
    direction $\mathbf{\hat{r}}$ and the direction to the observer, so
    $\phi=\pi/2+\theta$.
}
\end{figure}

Begin by calculating the surface brightness of a small segment of
length $\delta\ell_j$ in streamline $j$.
Summing the contributions from all such streamlines will
then provide the debris disk's total surface brightness along some line-of-sight.
Since these disks are often observed nearly edge on, an edge-on viewing geometry is
also assumed here, though these results are easily generalized for other
viewing geometries as well.
Segment $\delta\ell_j$ in streamline $j$ is composed of
dust having a total cross section $\delta\sigma_j$, so the
intensity of starlight $\delta I_j$ that is reflected by that segment is 
$\delta I_j=Q_{s}\Phi F_{i}\delta\sigma_j$, where
$F_{i}=L_\star/4\pi r^2$ is the flux of incident starlight,
$L_\star$ is the central star's luminosity, and
$r$ the dust grains' distance from the star. The dust grains'
phase function is $\Phi$, and it is normalized so that its
integral over all solid angles is unity, with this quantity
having units of steradians$^{-1}$. Note also that $F_{i}\delta\sigma$
is the power in the incident starlight while $\int\delta I_j d\Omega$
(when integrating over all solid angles $d\Omega$)
is the power in the scattered light,
so $Q_{s}=\int\delta I_j d\Omega/F_{i}\delta\sigma_j$ is the efficiency of
light scattering by these dust grains. In planetary astronomy
this quantity is known as the bond albedo \citep{LMT79}.

The surface brightness of the small segment
is $B_j=\delta I_j/\delta A_j$ where $\delta A_j$ is the projected
area on the sky occupied by that segment 
whose total length is $\delta\ell_j$ and projected length (as seen by the observer)
is $\delta x_j$; see Figure \ref{los_figure}.
Since that segment is actually a ribbon of material of height $2Ir$
due to the dust grain's inclinations $I$,
the segment's projected area is $\delta A_j=2Ir\delta x_j$,
and its surface brightness contribution is
$B_j=(Q_{s}\Phi F_{i}/2Ir)(\delta\sigma_j/\delta x_j)$.

The number of dust grains in segment $\delta\ell_j$ is
$\delta n_j=\lambda_j\delta\ell_j=\lambda_j v_j\delta t$
where $v_j$ is the dust grains' velocity
there and $\delta t$ is the time for the dust to traverse $\delta\ell_j$.
But this becomes $\delta n_j=(n_j/T_j)\delta t$ where $T_j$ is the dust grains'
orbit period, since $\lambda_j v_j=n_j/T_j$ where $n_j$ is the number of
dust grains in streamline $j$ that also have radius
$R_j=(\beta_{max}/\beta_j)R_{min}$. Consequently,
the cross section of dust in segment $\delta\ell_j$ is
$\delta\sigma_j=\pi n_j(\beta_{max}/\beta_j)^2R_{min}^2(\delta t/T_j)$.
That segment has length $\delta\ell_j=v_j\delta t$, and 
Figure \ref{los_figure} shows that its projected length is
$\delta x_j=v_j|\cos(\theta+\varphi)|\delta t$ where $\theta$ is the segment's
longitude as measured from the $\mathbf{\hat{x}}$ axis. The angle
$\varphi$ obeys $\sin\varphi=v_{j,\theta}/v_j$ and $\cos\varphi=v_{j,r}/v_j$
(see Figure \ref{los_figure}) 
where $v_{j,r}$ and $v_{j,\theta}$ are the dust grain's radial and tangential
velocities, Equations (\ref{velocities}), so
$\delta x_j = |v_{j,r}\cos\theta - v_{j,\theta}\sin\theta|\delta t=
2\pi a_j|\sin\theta + e_j\sin\tilde{\omega}_j|(\delta t/T_j)/\sqrt{1-e_j^2}$.
Inserting these results into $B_j$ and noting that
$n_j=N_j/N_l$ then yields
$B_j = B_0\sqrt{1-e_j^2}(\Phi\Omega_1N_j/N_l)
(R_{min}^2r_{out}^2/r^3a_j)(\beta_{max}/\beta_j)^2
/|\sin\theta + e_j\sin\tilde{\omega}_j|$
where the constant
\begin{eqnarray}
    \label{B0}
    B_0 &=& \frac{Q_sL_\star}{16\pi I r_{out}^2\Omega_1}.
\end{eqnarray}
Note that $\Omega_1=1$ steradian is introduced into the above so that
$B_0$ has units of surface brightness and that the combination $\Phi\Omega_1$
is dimensionless.
Summing the contributions from all streamlines then yields the disk's
total surface brightness $B(x)=\sum B_j$ as a function of projected distance
$x=r\cos\theta$ from the central star, so
\begin{eqnarray}
    \label{B}
    B(x) &=& B_0\sum_{i=1}^{N_{r\beta}}\sum_{j=1}^{N_l}\sum_{near}^{far}
        \frac{\Phi\Omega_1 N_i\sqrt{1-e_i^2}}{N_l|\sin\theta+e_j\sin\tilde{\omega}_j|}
        \left(\frac{R_{min}}{r_{out}}\right)^2\left(\frac{r_{out}^4}{r^3a_i}\right)
        \left(\frac{\beta_{max}}{\beta_i}\right)^2
\end{eqnarray}
where the innermost sum is over the two segments---one nearer and the
other further from the observer---that intersect the observer's line-of-sight
(see Figure \ref{los_figure}), while the middle sum proceeds over each streamlines'
orientation $\tilde{\omega}_j$, and the leftmost sum proceeds over the
$N_{r\beta}$ streamlines having distinct dust sizes $\beta_i$
that originate in the various planetesimal rings $r_i$. Equation (\ref{B})
also requires the two longitude $\theta(x)$ where the observer's line-of-sight
intercepts streamline $j$. That is obtained by
solving $x=r\cos\theta$ for $\theta(x)$ where
$r=p_j/[1 + e_j\cos(\theta - \tilde{\omega}_j)]$, which yields
$\theta(x)=\tilde{\omega}_j - \tan^{-1}(B/A) \pm \cos^{-1}(x/\sqrt{A^2+B^2})$
where $A=p_j\cos\tilde{\omega}_j - e_jx$, $B=p_j\sin\tilde{\omega}_j$,
and $p_j=a_j(1-e_j^2)$ is the ellipse's semi-latus rectum, with the $r$ in
Equation (\ref{B}) from $r=x/\cos\theta$.

The phase function employed here is the Henyey-Greenstein function
\begin{equation}
    \label{Phi}
    \Phi(\phi) = \frac{1-g^2}{4\pi(1+g^2-2g\cos\phi)^{3/2}}
        \quad\mbox{ster}^{-1}
\end{equation}
that is widely used in studies of circumstellar dust. The scattering angle
$\phi$ in the above is related to the dust grain's longitude $\theta$ via
$\phi=\pi/2 + \theta$; see Figure \ref{los_figure}.
This phase function is controlled by    
the dust grains' asymmetry parameter $g=\int\Phi(\phi)\cos\phi d\Omega$
(see Equation \ref{g}), with a positive value resulting in
forward light scattering while negative values result in backscattering.    

%Fig: surface brightness figure
\begin{figure}[t]
\epsscale{0.75}
\vspace*{-8ex}\plotone{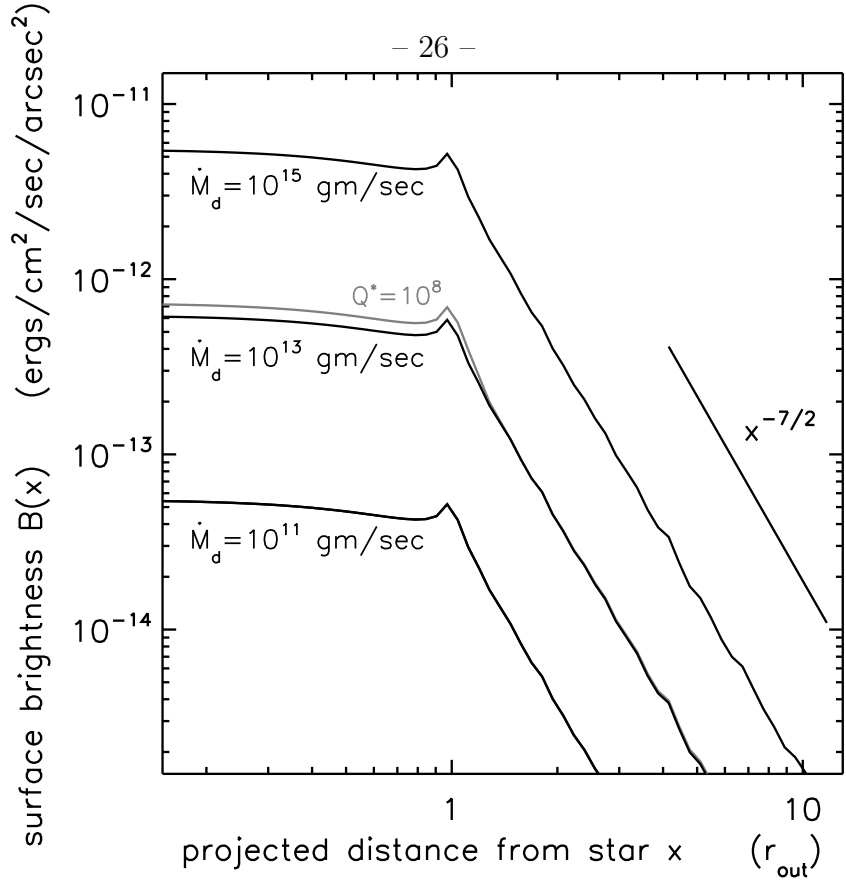}\vspace*{-3ex}
\figcaption{
    \label{sb_figure}
    Equation (\ref{B}) is used to calculate the
    surface brightness $B(x)$ versus projected distance from the star
    $x$ for three edge-on debris disks that are generated by narrow planetesimal rings
    that have the indicated dust production rates
    $\dot{M}_d=10^{11}$, $10^{13}$, and $10^{15}$ gm/sec.
    Black curves assume weak dust having $Q^\star<10^6$ ergs/gm,
    a light scattering asymmetry parameter $g=0.4$
    and albedo $Q_{s}=0.1$, with all
    system parameters identical to those of Figure \ref{n(tau)_figure}. 
    These curves give the disks' unfiltered surface brightnesses
    integrated over all optical wavelengths. A $x^{-7/2}$ curve is also shown,
    and the grey curve is for strong dust
    having $Q^\star=10^8$ ergs/gm.
}
\end{figure}

Equations (\ref{B}--\ref{Phi}) are used to calculate the surface brightness
for three edge-on debris disks whose parameters are identical to
those of Figure \ref{n(tau)_figure} except for differing dust production
rates $\dot{M}_d$; see Figure \ref{sb_figure}.
As expected, these surface brightness curves vary
as $B\propto\sqrt{\dot{M}_d}$. 
Also note that  $B(x)\propto x^{-7/2}$,
which again is in agreement with the type B debris disk of \cite{SC06}.

%Fig: surface brightness vs g
\begin{figure}[t]
\epsscale{0.75}
\vspace*{-8ex}\plotone{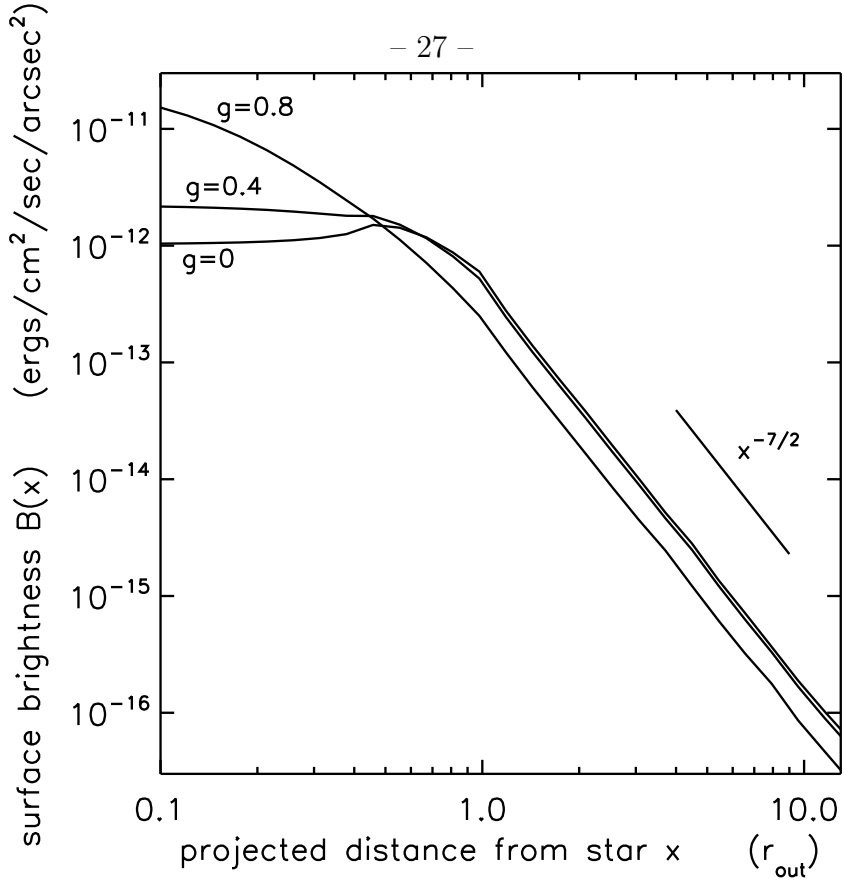}\vspace*{-3ex}
\figcaption{
    \label{sb_g_figure}
    Surface brightness $B(x)$ is plotted versus projected distance $x$
    for three edge-on debris disks
    that are generated by a broad ($r_{in}=0.5r_{out}$)
    planetesimal disk having a dust production rate $\dot{M}_d=10^{13}$ gm/sec,
    a light scattering efficiency of $Q_{scat}=0.1$, 
    and the indicated light scattering asymmetry parameter $g=0$, $0.4$,
    and $0.8$. Also shown is an $x^{-7/2}$ curve. 
    Note also that the small bump
    in $B(x)$ at $x=0.5r_{out}$ (which is where the line-of-sight
    runs along the planetesimal disk's inner edge) gets washed-out when
    the light scattering is very asymmetric with $g\gtrsim0.8$.
}
\end{figure}

Figure \ref{sb_g_figure} illustrates how the inner part
of a debris disk's surface brightness depends on the degree of asymmetry
in the dust grains' light scattering. 
For instance,
when the dust grains are either isotropic light scatters ($g=0$),
or are only modestly forward scattering ($g=0.4$), the debris
disk's surface brightness profile $B(x)$ is roughly constant
in the inner regions where $x\lesssim r_{out}$.
However, if the light scattering by the
dust is strongly asymmetric, with $g=0.8$, then $B(x)$ continues to increase
inwards of  $x=r_{out}$. Disks having this kind of knee-bend in their
surface brightness profiles appear to be common \citep{KAG05, GAK06}, 
with forward scattering of starlight being the preferred explanation
for the bent surface brightness profile of the edge-on debris
disk at AU Mic and $\beta$ Pic \citep{SC06, ACA09}.
Evidently, light scattering by
circumstellar dust is rather asymmetric.  
Lastly, note that almost identical results are obtained when the
dust grains are backscattering, with $g<0$, with those slight differences
being due to the dust grains' slightly larger radiation pressure efficiency
$Q_{rp}=1 - gQ_s$. And the following will assume that the grains are forward
scattering, which is appropriate for dust grains that are larger than
the wavelength of the incident radiation, and is consistent with 
forward scattering by  the Solar System's interplanetary dust \citep{LP86}.

\begin{figure}[t]
\epsscale{0.75}
\vspace*{-8ex}\plotone{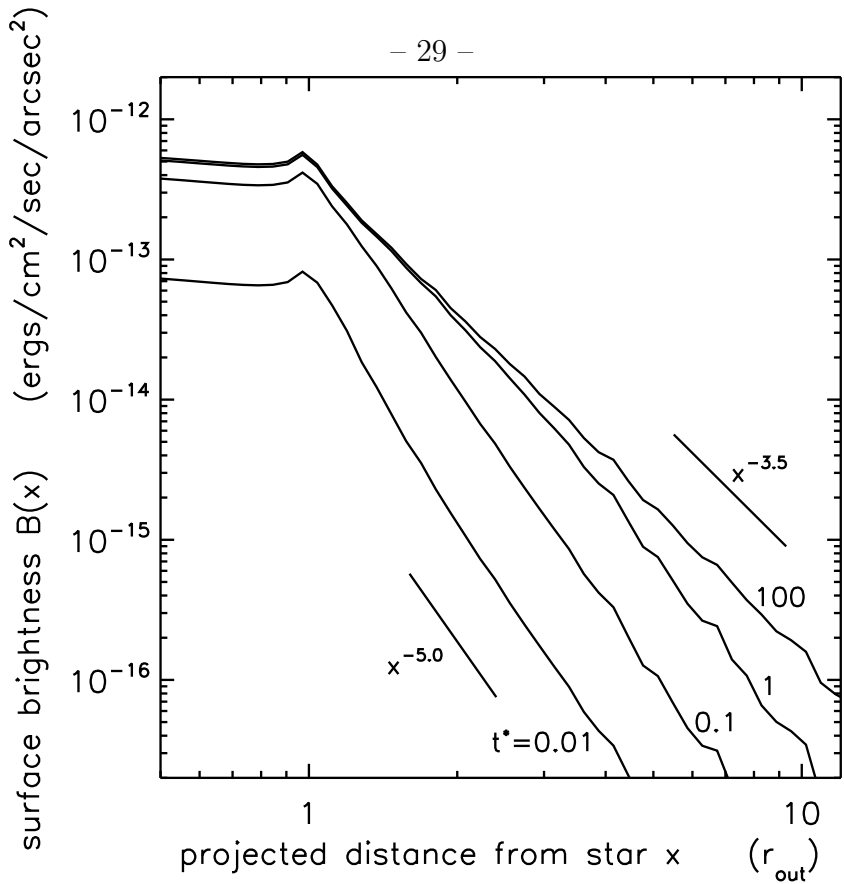}\vspace*{-3ex}
\figcaption{ 
    \label{sb_time_figure}
    Surface brightness $B(x)$ is plotted versus projected distance $x$
    (in units of $r_{out}$)
    for the debris-disk model that is described in Figures \ref{n(tau)_figure} and 
    \ref{tau_time_figure} viewed edge-on with $Q_s=0.1$. 
    Shown is $B(x)$ at dimensionless times
    $t^\star=t/T_0=0.01$, 0.1, 1, and 100, as well as power laws
    that vary as $x^{-3.5}$ and $x^{-5.0}$.
}
\end{figure}

Figure \ref{sb_time_figure} also shows how an edge-on disk's surface brightness
$B(x)$ evolves over time. As expected, $B(x)$ gets shallower at later times
as the smaller dust grains steadily populate the outer parts of the disk;
see also Figures \ref{n(tau)_figure} and \ref{tau_time_figure}. 
Eventually, the smaller grains arrive at collisional equilibrium at 
dimensionless times $t^\star=t/T_0\gg1$, and the disk surface
brightness settles into the expected $B(x)\propto x^{-3.5}$ power law.
But note that at earlier times, such as at time $t^\star=0.01$,
the disk's surface brightness can
be as steep as $B(x)\propto x^{-5.0}$. Consequently, disks having
a surface brightness steeper than $x^{-3.5}$ might indicate
that the disk is younger than time $T_0$, or
that dust production by the planetesimal
disk has increased in recent times.

%Fig: surface brightness vs time

Note that these surface brightness
calculations assume that the dusty disk is so tenuous
that the grains do not shadow each other; this assumption is confirmed in
Section \ref{shadowing}.
Also keep in mind that Figures \ref{sb_figure}--\ref{sb_time_figure}
give the disks' total surface brightness integrated over all optical
wavelengths. If these systems were instead observed through a narrowband
filter having a central frequency $\nu$ 
then the star's total luminosity $L_\star=
4\pi^2R_\star^2\int_0^\infty B_\nu d\nu = 4\pi R_\star^2\sigma T_\star^4$
(where $R_\star$ is the stellar radius, $B_\nu$ is the Planck
function, $\sigma$ the Stefan-Boltzmann constant, and
$T_\star$ is the star's effective temperature)
in Equation (\ref{B0}) should 
instead be replaced by its specific luminosity
$\partial L_\star/\partial\nu = \pi L_\star B_\nu/\sigma T_\star^4$,
so that Equations (\ref{B}--\ref{Phi}) would then provide the disk's surface
brightness per unit frequency.

%%%%%%%%%%%%%%%%%%%%%%%%%%%%%%%%%%%
% Stopping dust production 
%%%%%%%%%%%%%%%%%%%%%%%%%%%%%%%%%%%
\subsubsection{a relic debris disk}
\label{relic}

It is also worth considering the debris disk's evolution when the
planetesimal disk's dust production suddenly ceases. Such might
occur if there are any planets in the system that quickly adjust their orbits,
perhaps due to a rapid migration of a planet through a dense planetesimal disk
\citep{GML04}, or due to planet-planet scattering
({\it i.e.}, the Nice model, \citealt{GLT05}).
Should that occur, then the planetesimals might find themselves in
unstable orbits, which then leads to a rapid dynamical erosion of the disk
and a cessation of dust production.

\begin{figure}[t]
\epsscale{0.75}
\vspace*{-8ex}\plotone{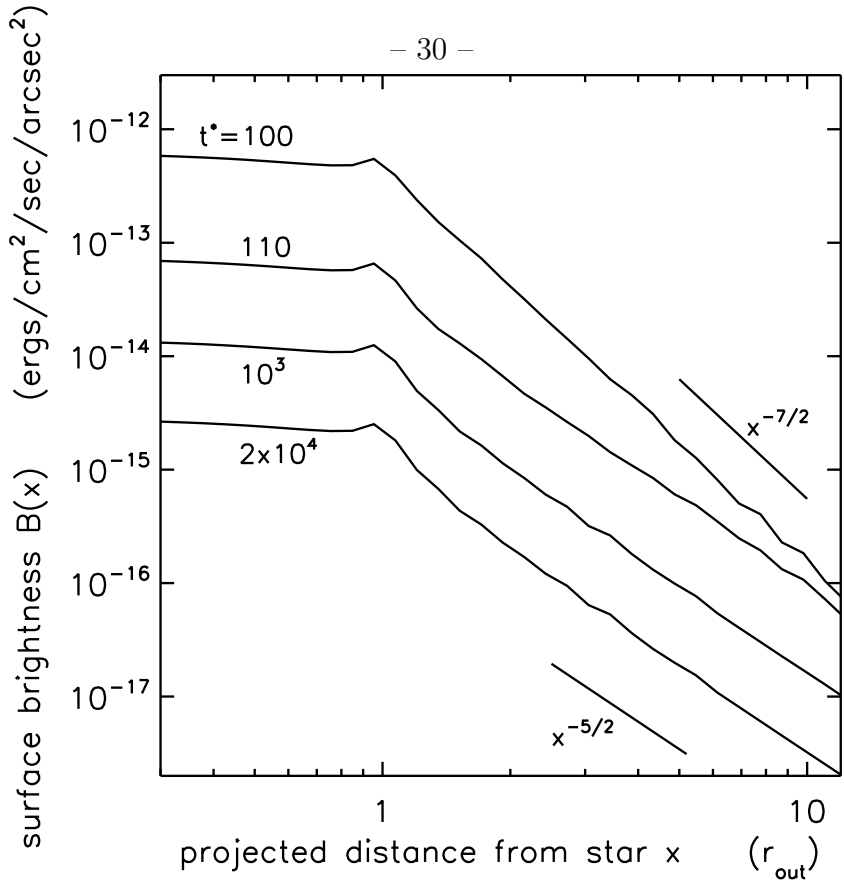}\vspace*{-3ex}
\figcaption{ 
    \label{relic_figure}
    The dust-producing debris disk model of Figure \ref{n(tau)_figure} is evolved until
    dimensionless time $t^\star_{stop}=100$ when dust production ceases
    ($P_i=0$). The debris disk's surface brightness $B(x)$ is shown
    at the indicated times $t^\star$, as well as two power laws
    $x^{-7/2}$ and $x^{-5/2}$.
}
\end{figure}

Halting dust production is simulated here
by setting the planetesimal ring's dust production rate $P_i(t)=0$ at all
times after $t_{stop}$, or equivalently changing the $1\rightarrow0$
in the invariant evolutions equation (\ref{dn/dt}) at dimensionless times
$t^\star>t^\star_{stop}$ where $t^\star_{stop}=t_{stop}/T_0$.
Solving those equations numerically shows that the larger grains in the
inner part of the debris disk are quickly destroyed due to their
short collisional lifetimes (Figure \ref{T_col_figure}). This depletes
the inner disk's optical depth and drives it towards a
shallower $\tau_n(r)\propto r^{-1/2}$
power law that decreases over time. And as Figure \ref{relic_figure} shows,
the cessation of dust production
also decreases the inner part of an edge-on disk's surface brightness
so that $B(x)\propto x^{-5/2}$ at times $t^\star \gg t^\star_{stop}$.
So the detection of a debris disk having a shallow optical
depth profile $\tau_n(r)\propto r^{-1/2}$ (when the view to the disk
is oblique or face-on), or an edge-on disk having a shallow 
$B(x)\propto x^{-5/2}$ surface brightness profile,
would indicate that the object is a {\em relic} disk
wherein dust-production has ceased. That relic disk will be
composed mostly of small, marginally-bound grains of radii $R\simeq R_{min}$.
The abundance $N_S$ of those small dust grains will then fade 
over time according to Equation (\ref{dN/dt_singlesize}),
whose solution is
$N_S(t')=N_S(t_{stop})/[1 + N_S(t_{stop})\bar{\alpha}_{SS}t'/T_{out}]$
when the dust production rate $P_S=0$,
where $t'=t-t_{stop}$ is the time since the end of dust production
and $\bar{\alpha}_{SS}$ is the small grains' collisional probability density.
Note that this behavior over time is typical for a system that
steadily grinds away without any replenishment ({\it cf.}, \citealt{WSG07, LKR08}).
Inspection of the upper two curves in Figure \ref{relic_figure} also
shows that the time for the system to transform into a relic  disk
is about $t_{trans}\sim10T_0\sim0.1T_{eq}$ {\it i.e.},
the transition occurs relatively quickly in comparison
to the system's equilibrium timescale $T_{eq}$.

%Fig: relic disk

%%%%%%%%%%%%%%%%%%%%%%%%%%%%%%%%%%%
% beta Pic
%%%%%%%%%%%%%%%%%%%%%%%%%%%%%%%%%%%
\section{Application to $\beta$ Pictoris}
\label{beta_pic}

\begin{figure}
\epsscale{0.75}
\vspace*{-6ex}\plotone{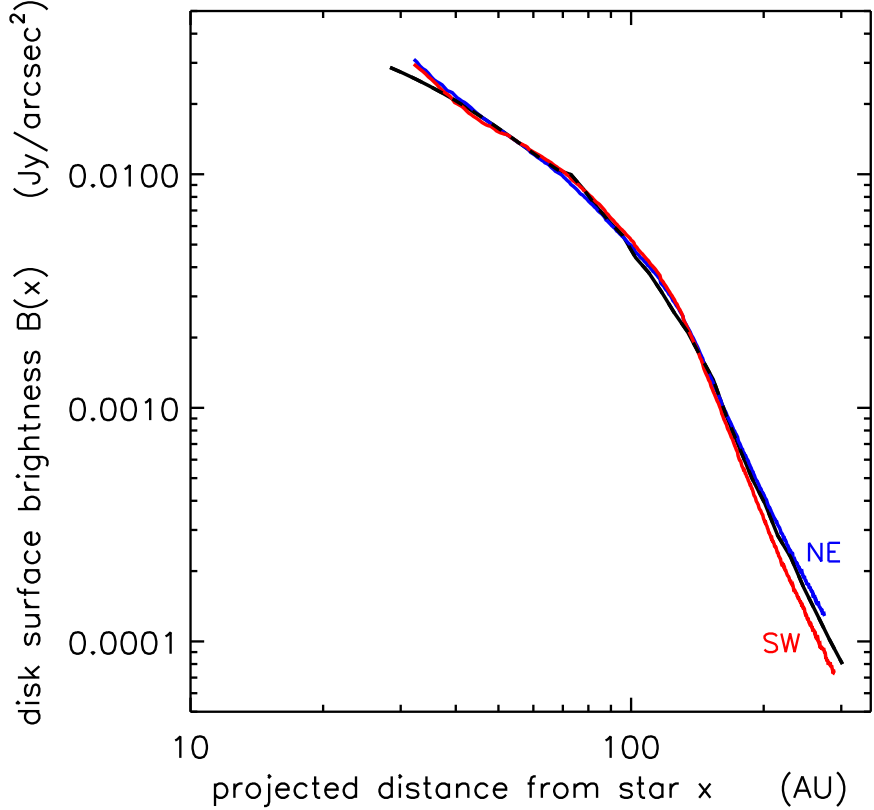}\vspace*{-3ex}
\figcaption{
    \label{Bpic_figure}
    Red and blue curves give
    the surface brightness $B(x)$ of the edge-on disk orbiting $\beta$ Pictoris,
    extracted from the Hubble Space Telescope observations acquired by \cite{GAK06}
    in the F606W filter, and plotted versus projected
    distance $x$ from the central star. The black dashed curve is for a 
    model that adopts a $q=2.5$ size distribution for dust
    produced by a broad disk of planetesimals
    orbiting at $75<r<150$ AU from the central star; this system was
    evolved for $t=1.2\times10^7$ yrs, which is the age of the central star.
    The simulated dust grains' inclinations are
    $I=0.07\mbox{ rad}=4.0^\circ$, their density is $\rho=1$ gm/cm$^3$,
    and they are strong with $Q^\star=10^8$ ergs/gm. The dust grains also
    have a light scattering efficiency $Q_s=0.7$,
    a light-scattering asymmetry parameter $g=0.67$,
    and a radiation pressure efficiency $Q_{rp} = 1 - gQ_s=0.53$.
    The planetesimal disk's dust production rate is
    $\dot{M}_d=1.7\times10^{15}$ gm/sec, so $N_0=3.6\times10^{34}$ and
    $T_0=1.5\times10^5$ yrs, and the disk's
    equilibrium timescale is $T_{eq}=2.0\times10^7$ yrs.
}
\end{figure}

The colored curves in Figure \ref{Bpic_figure} show
the surface brightness of the starlight that
is scattered by the edge-on debris disk orbiting $\beta$ Pictoris.
These curves are extracted from images that \cite{GAK06}
acquired with the Hubble Space Telescope (HST) at optical wavelengths.
These profiles exhibit the classic signature of a
debris disk that is generated by a disk
of colliding planetesimals that extends out to about 
$r_{out}\simeq150$ AU, with the surface brightness
falling off steeply as $B(x)\propto x^{-3.5}$
at projected distances $x\gtrsim220$ AU, and less steep interior to $r_{out}$.
Note also that the $\beta$ Pic disk is asymmetric, with the outer part of the
disk's northeast (NE, blue curve) ansa being about $50\%$ brighter than its
southwest (SW, red curve) ansa. Although that asymmetry is not accounted for
by the axially symmetric debris-disk model that is used here,
possible causes for that
asymmetry are described in Section \ref{disk_asymmetry}.

%Fig: beta pic model

$\beta$ Pic is an A5V star that
lies a distance $\Delta=19.28$ pc away, has a luminosity $L_\star=8.7$L$_\odot$,
mass $M_\star\simeq1.8$M$_\odot$, an effective temperature $T=8200$ K
\citep{CVL97}, and an age of $t_\star\sim12$ Myrs \citep{ZSB01}.
Note that the star's peak emission occurs at wavelength
$\lambda=0.35\ \mu$m, yet the radius of the smallest bound grain
is $R_{min}\simeq3\ \mu$m assuming the grains have a density of $\rho=1$ gm/cm$^3$
and a radiation pressure efficiency $Q_{rp}=0.53$ (Equation \ref{R_min}).
So this system is in the geometric optics limit, and the light-scattering
theory employed here is appropriate. Figure 11 of \cite{GAK06}
also shows that the disk's vertical distribution has a full width at half
maximum (FWHM) that varies as $\mbox{FWHM}\propto0.14x\sim2h$, so the dust
grains' inclinations are approximately $I=h/x\sim0.07$ radians. 

The black curve in Figure \ref{Bpic_figure} shows that the debris disk model
can reproduce $\beta$ Pic's observed surface brightness when
the dust-producing planetesimal disk is quite broad, with inner and outer
radii $r_{in}=75$ AU and $r_{out}=150$ AU. This system has been evolved
over the star's lifetime, $t_\star=12$ Myrs, so the system is still
approaching equilibrium since $T_{eq}=20$ Myrs.
The dust grains must also be rather asymmetric light scatters,
with $g=0.67$ (assuming forward scattering),
in order to account for the knee seen in the
surface brightness profile at $x\sim150$ AU, consistent with
\cite{ACA09}. The model also assumes that the
dust grains are rather reflective, $Q_s=0.7$, which is the Bond albedo
of Saturn's icy A and B rings at the observation wavelength \citep{PBB05}.
With these assumption in hand, fitting the simulated
surface brightness to the disk's observed $B(x)$ requires the
planetesimal disk's dust production rate to be
$\dot{M}_d=1.7\times10^{15}$ gm/sec.
The total cross section of dust in the simulated debris disk is
$A_d=1.9\times10^{20}$ km$^2$, and the total 
mass of dust is $M_d=11$ lunar masses. Note that this mass
is comparable to the 8 lunar masses
that \cite{HGZ98} inferred from  submillimeter observations
of this disk, which supports the contention
that the $\beta$ Pic dust grains are rather reflective. 

The dust production rate inferred here is about $\dot{M}_d\simeq9$ M$_\oplus$/Myr.
For comparison, this rate is about 200 times higher
than what \cite{SC97} report for their model of the collisional
erosion of our own Kuiper Belt. Note that $\beta$ Pic's
dust production rate is quite considerable, for 
if it has held steady over the system's lifetime, then the unseen planetesimals
orbiting $\beta$ Pic would have lost a total mass of
$\dot{M}_dt_\star\sim110$ earth-masses.

It should be noted that the planetesimal disk's inferred erosion rate
is very sensitive to the dust grain's optical properties. Equation (\ref{B}) shows
that an edge-on disk's surface brightness is varies as
$B\propto B_0N_0 R_{min}^2$ where $B_0\propto Q_s$,
$N_0\propto \dot{M}_d^{1/2}\rho^2Q_{rp}^{-5/2}$, and
$R_{min}\propto Q_{rp}/\rho$ (Equations \ref{Q_rp0}, \ref{R_min}, \ref{N03},
and \ref{B0}),  which means that the inferred
dust production rate varies as $\dot{M}_d\propto f_Q$
where $f_Q=Q_{rp}/Q_s^2$ and $Q_{rp}=1-gQ_s$.
(Interestingly, the inferred dust production rate is insensitive
to the assumed grain density $\rho$, which cancels out.)
So if the dust grains are instead dark and have a scattering efficiency
$Q_s=0.1$, then the radiation pressure efficiency is
$Q_{rp}\simeq1$ and $f_Q\simeq100$, which means that the
dust production rate must be $\sim100$ faster if the dust grains
are dark. In this case, $\beta$ Pic would have lost
$\sim\dot{M}_dt_\star\sim10^4$ earth-masses
over its lifetime, which is implausible. Rather, it is more likely that
the dust grains at $\beta$ Pic are bright.

%Fig: beta pic optical depth
\begin{figure}[t]
\epsscale{0.75}
\vspace*{-6ex}\plotone{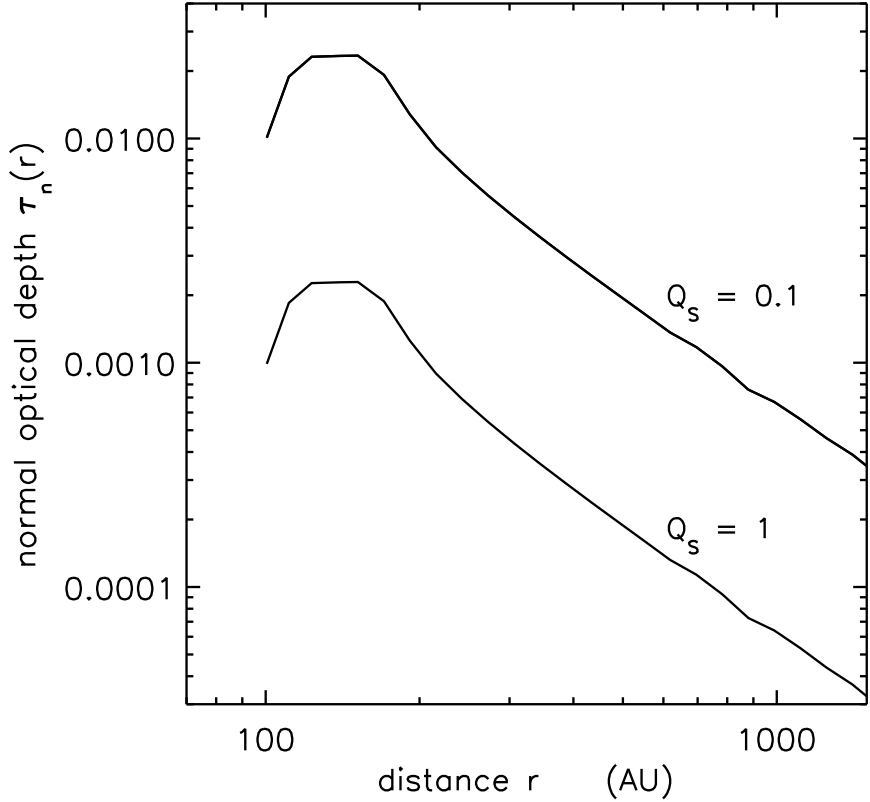}\vspace*{-3ex}
\figcaption{
    \label{Bpic_tau_figure}
    Normal optical depth $\tau_n(r)$ that results in the
    surface brightness profile shown in Figure \ref{Bpic_figure}, assuming
    $Q_s=0.1$ and $Q_s=1$. The dust production rate for the 
    $Q_s=1$ model is 300 times smaller than the $Q_s=0.1$ model.
}
\end{figure}

The optical depth profile $\tau_n(r)$ that is inferred from $\beta$ Pic's observed
surface brightness $B(x)$ is shown in Figure \ref{Bpic_tau_figure}. Note that
$B(x)\propto \int Q_s\tau_n d\ell$ where the integration is along the line of sight,
so the disk's optical depth is uncertain by a factor
of $Q_s$. The $Q_s=1$ curve obtained here is similar
to the optical depth profile that \cite{ACA09} inferred from the same
dataset\footnote{The \cite{ACA09} work does not mention $Q_s$;
we suspect that $Q_s=1$ was assumed there.}. The range of possible
values for the total dust mass $M_d$ and total dust cross section $A_d$
are $4.7\le M_d\le 170$ lunar masses and
$1.2\times10^{20} \le A_d\le 1.6\times10^{21}$ km$^2$,
with the lower values for a disk composed of bright $Q_s=1$ dust grains,
and the higher values for dark $Q_s=0.1$ dust.

%%%%%%%%%%%%%%%%%%%%%%%%%%%%%%%%%%%
% Sensitivity
%%%%%%%%%%%%%%%%%%%%%%%%%%%%%%%%%%%
\subsection{sensitivity to parameters}
\label{sensitivity}

The following explores the sensitivity of the above results for $\beta$ Pic
to the assumed model parameters.

First, the dust produced by the simulated planetesimal disk has a
$q=2.5$ size distribution, so most of the just-produced dust grains'
mass is in the largest grains that actually
contribute little to the debris disk's optical depth. Consequently, the
dust production rate $\dot{M}_d$ inferred above is somewhat sensitive to
$\beta_{min}$, which is the size parameter of the largest grains that are allowed
in the model. The equilibrium model shown in Figure \ref{Bpic_figure}
has $\beta_{min}=0.065$ (which corresponds to a
maximum grain radius $R_{max}=23\ \mu$m),
and it requires a dust production rate of 
$\dot{M}_d=1.7\times10^{15}$ gm/sec when the model is fit to the
the disk's observed surface brightness profile assuming $Q_s=0.7$.
However, cutting off the dust size distribution at $\beta_{min}=0.1$ and then
optimizing the other parameters then results in a fit that is only marginally 
acceptable; that model has $R_{max}=14\ \mu$m
and $\dot{M}_d=1.2\times10^{15}$ gm/sec. Nonetheless, models having
$\beta_{min}\gtrsim0.2$ are ruled out, because they are deficient in
grains having sizes $R>6\ \mu$m that would be confined to radial distances
of $75\lesssim r \lesssim250$ AU, and their absence from the model
reduces the simulated disk's surface brightness at projected distances
$x\sim r$. 

Interestingly, the disk's surface brightness profile $B(x)$ is rather sensitive
to the size distribution power law $q$. For instance, models having $q\le2.0$
do not reproduce the disk's observed surface density profile,
because the model disk's inner $r\lesssim150$ AU region
is overdense with large dust grains that contribute too much surface brightness
at $x\lesssim150$ AU, while models having $q\ge3.0$ are underdense in large grains
and thus too dim there. Evidently, the dust being produced by the unseen
planetesimals at $\beta$ Pic have a $q\simeq 2.5$ size distribution 
that is just a bit shallower than the canonical  Dohnanyi $q=3.5$ size distribution.
And because the observed dust size distribution is constrained to
$2<q<3$, this translates into an uncertainty in the value of
$\dot{M}_d$ quoted above is no more than a factor of 2.
Also, do keep in mind that the value for
$q$ refers to the size distribution for dust production by the planetesimal
disk, and that the debris disk's resulting dust size distribution is 
actually much steeper
due to collisions among dust grains (e.g., Figure \ref{N_final_figure}).

This kind of modeling is also sensitive to the planetesimal disk's 
inner and outer radii. The midpoint of the knee  at $x\simeq110$ AU in 
$\beta$ Pic's surface brightness profile  indicates 
that the planetesimal disk's midpoint is near
$r\simeq110$ AU (Figure \ref{Bpic_figure}),
and best agreement with the observations is
achieved when $r_{in}\simeq75$ AU and $r_{out}\simeq150$ AU {\it i.e.},
when the disk has a radial aspect ratio of $r_{in}/r_{out}\simeq0.5$.
However, broader disks having $r_{in}/r_{out}<0.3$ are ruled out
because the knee in their surface brightness profiles are too broad to
fit the observations. Likewise,
narrower planetesimal disks having $r_{in}/r_{out}\ge0.8$
are ruled out because those models produce a prominent bump
in the debris disk's surface brightness where $x=r_{in}$,
and such bumps are not seen in these optical observations.
Although these surface brightness bumps are seen in infrared observations
of $\beta$ Pic \citep{WKR05, TFW05},
their locations vary with the observing wavelength, which
indicates that their interpretation likely requires modeling the disk's
thermal emission as well.

We have also considered `inside-out' erosion
of the planetesimal disk, with $c=-2$, but the edge-on disk's surface brightness was
rather similar to that generated by planetesimal disk suffering `outside-in' erosion
with $c=+2$, so $c=0$ is adopted in all models shown here.
Also note that strong dust grains having $Q^\star=10^8$ ergs/gm are
preferred over models that use weaker $Q^\star\le10^7$ ergs/gm dust grains.
The larger grain strength increases the survival of the
larger dust grains (see Figure \ref{N_final_figure}) that are confined
to the vicinity of the planetesimal disk at $75<r<150$ AU. 
When $Q^\star\le10^7$ ergs/gm, the lower abundance of large dust grains then
results in a noticeable surface brightness deficit at $x\sim100$ AU.

%%%%%%%%%%%%%%%%%%%%%%%%%%%%%%%%%%%
% implications
%%%%%%%%%%%%%%%%%%%%%%%%%%%%%%%%%%%
\subsection{implications for planet formation}
\label{implications}

If one adopts the plausible assumption
that the $\beta$ Pic dust grains are bright ($Q_s\sim0.7$), then the
mass of the planetesimal disk is likely well above 110 M$_\oplus$, which is
sufficient to form about six Neptune-class planets. This
planetesimal disk's considerable mass, plus its broad radial extent
($75\lesssim r<150$ AU), also provides several interesting constraints on the
prospects for planet formation at $\beta$ Pic.
Analogy with the Kuiper Belt, which resides at distances $\sim30\%$ beyond Neptune's
orbit, suggests that there should be no Neptune-class giant planets orbiting
well beyond $r\gtrsim60$ AU from $\beta$ Pic, because such planets would
have scattered away a portion of the inferred planetesimal disk
and part of its dust disk.
The fact that there are no such planets yet
means that if planets are trying to form there via
core-accretion (e.g., a runaway accretion of planetesimals by a
growing planetary embryo), then the timescale for this kind of assembly in the
$75\lesssim r<150$ AU zone must be longer than the system's
$t_\star\sim12$ Myr age. However, gravitational instability (GI)
is a much faster planet-formation mechanism, one that can also 
form giant planets at large stellarcentric distances \citep{DBM07, DVF09}.
However, GIs are only operative when the circumstellar disk
is still rich in nebula gas, and the lifetime of a gas disk
is at most a few Myrs. Consequently, the presence of
$\beta$ Pic's broad and massive
planetesimal disk also tells us that planet formation via GI did not
happen in the $75\lesssim r<150$ AU zone.

It can also be concluded that giant planets are unlikely to have formed interior
to $r\sim75$ AU and subsequently migrated or scattered deep into the 
$75\lesssim r<150$ AU zone. For if that were to have happened at
$\beta$ Pic, then the migrating/scattering planet
would have accreted and/or scattered the planetesimal disk,
causing the circumstellar dust to transition into a relic
debris disk within a time $t_{trans}\sim0.1T_{eq}\sim2$ Myrs.
Consequently, the probability
of an astronomer observing this system before its surface brightness
transitioned into a relic $B(x)\propto x^{-5/2}$ profile  
is $t_{trans}/t_\star\sim15\%$. This finding is also consistent with
\cite{BWM09}, who show that zero of 106 stars having circumstellar dust
also appear to have experienced  a Nice-model scattering of giant planets.

Of course, the Nice model was not intended to
explain observations of circumstellar debris disks. However, our purpose here
is to show how disk observations can be used to determine whether
any particular Solar System's scenario might
also be ongoing at other circumstellar disks, or if that scenario should
be regarded as a rarity among the known population of star-disk systems.
Indeed, it is a curiosity that $\beta$ Pic seems to be planetless at
$r\gtrsim75$ AU zone, despite having a substantial amount of
planetesimal mass there. However $\beta$ Pic's heavy dust production rate,
which is at least $\dot{M}_d\sim9$ M$_\oplus$/Myr (and may be much larger
if the grains are darker than $Q_s=0.7$), suggests an alternate interpretation---that
the planetesimal disk orbiting $\beta$ Pic is not a
planet-producing disk, but is instead a
planetesimal-destroying disk, due to a vigorous collisional grinding
of planetesimals and a heavy mass-loss rate due to the
blowout of dust by radiation pressure.

%%%%%%%%%%%%%%%%%%%%%%%%%%%%%%%%%%%
% Assumptions
%%%%%%%%%%%%%%%%%%%%%%%%%%%%%%%%%%%
\subsection{Testing the assumptions}
\label{assumptions}

This subsection examines the key assumptions that are employed here.
Those assumptions are: (1) that collision fragments contribute little to the
disk's optical depth, (2) the dusty disk is so tenuous
that the dust grains are not shadowing each other,
(3) that Poynting Robertson drag is insignificant, and
(4) that the planetesimal disk's dust production is steady over time.
A final subsection then comments on the fact that many debris disks
are non-axisymmetric.

%%%%%%%%%%%%%%%%%%%%%%%%%%%%%%%%%%%
% Fragmentation
%%%%%%%%%%%%%%%%%%%%%%%%%%%%%%%%%%%
\subsubsection{dust fragmentation}
\label{fragmentation}

When dust grains collide, they spawn smaller dust fragments,
but their contribution to the debris disk's optical depth is ignored by
this model. This is appropriate if the bulk of those fragments are sufficiently
small and fast such that they are unbound and  leave the system.
It turns out that this assumption is quite reliable, and is confirmed
from the model output. Begin with a single term in Equation (\ref{dN^c}),
${\cal R}_{ijk}=\alpha_{ijk}n_in_j/T_{out}$,
which is the rate at which the dust in the target streamline $i$ are destroyed
due to collisions with the dust in streamline $j$ at the site $r=r_{ijk}$
where their orbits intersect. The resulting dust fragments
are presumed to have a power-law
size distribution $N_f=CR_f^{-q}$, which
gives the number of dust fragments of radius $R_f$ produced by the
destruction of a single target grain of radius $R_i$, with the coefficient $C$
determined by mass conservation. Equations (\ref{v_i'}--\ref{v_recoil}) then
provide the speed $v_f$ at which the fragments recoil from the collision,
and their specific energy $E=\onehalf v_f^2 - GM_\star(1-\beta_f)/r_{ijk}$
indicates whether the collision fragments having a size parameter $\beta_f$
will stay bound to the star or  leave the system. The total rate at which
the debris disk produces collision fragments of radius $R_f$
is $\tilde{{\cal R}}_f=\sum N_f{\cal R}_{ijk}$ where the sum proceeds over
all sites in the disk $r_{ijk}$ where dust grains collide and produce bound
fragments of radii $R_f$. This is the secondary dust production rate,
and it is to be compared to the planetesimal disk's dust production
rate $P_f$. The ratio $\tilde{{\cal R}}_f/P_f$
depends of the radius of the smallest
possible fragment $R_s$. Also recall that $R_{min}$ is the radius of the smallest
bound dust grain that is produced in the planetesimal disk, so $R_s<R_{min}$.
In this instance,
the above algorithm finds that $\tilde{{\cal R}}_f\ll P_f$, which means that
the rate at which collision are injecting second-generation dust fragments
into the debris disk is negligible in comparison to the rate at which the
planetesimal disk is injecting first-generation dust into the disk.
For example, when $R_s<0.5R_{min}$, this algorithm
finds that $\tilde{{\cal R}}_f/P_f<1\%$ when applied to
the example model of Section \ref{example}. The ratio $\tilde{{\cal R}}_f/P_f$
is small because most of the collision fragments are small, unbound,
and do not contribute to the debris disk, which justifies
the model's neglect of dust fragmentation.

%%%%%%%%%%%%%%%%%%%%%%%%%%%%%%%%%%%
% Shadowing
%%%%%%%%%%%%%%%%%%%%%%%%%%%%%%%%%%%
\subsubsection{shadowing}
\label{shadowing}

If the disk is sufficiently dense, then dust grains can shadow each
other and alter the disk's surface brightness profile $B(x)$.
The calculations presented above all assumed that the dust grains are not shadowing
each other, and that assumption is confirmed by examining the disk's
radial optical depth $\tau_r$ \citep{W10}. Begin with
 a narrow annulus in the disk
of radius $r$ and radial width $\Delta r$. The cross section of dust in that
annulus is $\Delta A=\tau_n(r)2\pi r\Delta r$ where $\tau_n(r)$ is the disk's
normal optical depth. The annulus' vertical thickness is $2Ir$, so its
optical depth along a radial line-of-sight is
$\Delta \tau_r=\Delta A/2\pi r2Ir=\tau_n\Delta r/2Ir$.
The total radial optical depth of dust interior to $r$ is then
\begin{equation}
    \label{tau_r}
    \tau_r(r) = \int_0^r\frac{\tau_n(r')dr'}{2Ir'},
\end{equation}
which is easily integrated numerically. When this quantity is not small,
then the dust grains at $r$ are shadowed by interior dust, which reduces the
stellar illumination there by a factor $e^{-\tau_r}$.
But if $\tau_r\ll1$ then the dust grains are fully illuminated.

Most of the debris disk models described in the previous sections have
$\tau_r<10^{-3}$. The two exceptions are the heavy dust-producing models,
such as those having $\dot{M}_d=10^{15}$ gm/sec that have
$\tau_r\sim0.01$ (Figures \ref{T_col_figure},
\ref{tau_figure}, and \ref{sb_figure}),
while the $\beta$ Pic model that considers
dark $Q_s=0.1$ dust grains has $\tau_r\sim0.04$. So in summary, shadowing
by dust grains is not significant for the debris-disk models considered here.

%%%%%%%%%%%%%%%%%%%%%%%%%%%%%%%%%%%
% PR drag
%%%%%%%%%%%%%%%%%%%%%%%%%%%%%%%%%%%
\subsubsection{Poynting Robertson drag}
\label{pr_drag}

Poynting-Robertson (PR) drag is a weak force that causes the orbits of
small dust grains to slowly decay. The following shows that this orbital decay is
negligible for the dust in the $\beta$ Pic debris disk.

PR drag is the acceleration that results when a moving grain absorbs and/or
scatters stellar photons. Because a grain's motion also results in
a slight `headwind' of photons, the transfer of momentum, from the photons to
the grain, causes its orbit to decay. The acceleration on a dust grain due to
PR drag is \citep{BLS79}
\begin{equation}
    \label{a_pr}
    \mathbf{a}_{PR} = -\frac{\beta GM_\star}{r^2}\left(\frac{2v_r}{c}\mathbf{\hat{r}}
        + \frac{v_\theta}{c}\mathbf{\hat{\theta}}\right)
\end{equation}
where $v_r$ and $v_\theta$ are the grain's radial and tangential velocities
(Equations \ref{velocities}) and $\mathbf{\hat{r}}$, $\mathbf{\hat{\theta}}$
are unit vectors in a polar coordinate system.
Inserting this into the Lagrange planetary equations and
time-averaging those equations over an orbit then provides the
rates at which an orbiting dust grain's semimajor axis $a$ and eccentricity $e$
decay due to PR drag,
\begin{mathletters}
    \begin{eqnarray}
        \dot{a} &=& -\frac{2\beta GM_\star}{ac}\frac{1+\frac{3}{2}e^2}{(1-e^2)^{3/2}}\\
        \mbox{and}\quad \dot{e} &=& -\frac{5\beta GM_\star e}{2\sqrt{1-e^2}a^2c}
    \end{eqnarray}
\end{mathletters}
\citep{WW50, BLS79}. The timescales associated with this orbit decay are
\begin{mathletters}
    \label{T_PR}
    \begin{eqnarray}
        T_a &=& \frac{a}{|\dot{a}|} =
             \frac{(1-e^2)^{3/2}}{\beta(1+\frac{3}{2}e^2)}
             \left(\frac{a}{r_{out}}\right)^2T_{PR}\\
        \mbox{and}\quad T_e &=& \frac{e}{|\dot{e}|} = 
            \frac{4\sqrt{1-e^2}}{5\beta}\left(\frac{a}{r_{out}}\right)^2T_{PR}
    \end{eqnarray}
\end{mathletters}
where the constant
\begin{equation}
    T_{PR} = \frac{r_{out}^2c}{2GM_\star} = 
        2.0\times10^6\left(\frac{M_\star}{M_\odot}\right)^{-1}
        \left(\frac{r_{out}}{50\mbox{ AU}}\right)^{2}\mbox{ yrs}.
\end{equation}
Inserting Equations (\ref{a}) and (\ref{e}) into the above shows
that these timescales are simple functions of grain size $\beta$ or $R$.
If these orbit decay timescales are long compared to
a dust grains' collisional lifetime $T_c$, then it is appropriate to ignore PR
drag.

These orbit decay timescale are plotted in Figure \ref{T_col_figure}
for the example debris disk that is generated by a
planetesimal ring of radius $r_{out}=50$ AU (see
Sections \ref{example} --\ref{brightness}
and Figures \ref{n(tau)_figure}--\ref{sb_time_figure}). All of
these curves diverge for small grains that have radii $R\simeq R_{min}$
and size parameters $\beta\simeq\onehalf$. Note that for most grain sizes
$T_c(R)\ll T_a(R)$, which means that the effects of PR drag are negligible over
the lifetime of grains that are only slightly larger than $R_{min}$.
Figure \ref{T_col_figure} also
shows shows that PR drag is only significant when the system's 
dust production rate $\dot{M}_d$ is sufficiently low, which can make
$T_c\gtrsim T_a$ large, but only for these small $R\simeq R_{min}$ grains.
But this only occurs when $\dot{M}_d\lesssim10^{11}$ gm/sec,
for the scenario considered in Figure \ref{T_col_figure}.
A similar result was obtained by \cite{W05}, who showed that PR
drag is only significant when the debris disk is sufficiently tenuous. 
And finally, note that PR drag is completely negligible ($T_c\ll T_a$)
for grains of all sizes in the $\beta$ Pic debris disk, due to its very
vigorous dust production rate and the grains' short collisional lifetimes.

%%%%%%%%%%%%%%%%%%%%%%%%%%%%%%%%%%%
% Fragmentation
%%%%%%%%%%%%%%%%%%%%%%%%%%%%%%%%%%%
\subsubsection{dust production over time}
\label{dust_production}

The model employed here assumes that the planetesimal
disk's dust production rate is steady over time. But this assumption
might seem debatable, because the collisional erosion that drives
dust production will ultimately decrease the planetesimal disk's mass over time.
But keep in mind that accretion within the planetesimal disk tends to
produce larger bodies whose gravity can stir-up the planetesimal disk,
and that can instead increase the disk's dust production rate. Further, 
simulations of the collisional/accretional evolution of a planetesimal disk
shows that the outcomes are very sensitive to the planetesimal disk's
initial conditions that are poorly known \citep{W10}.
Due to this uncertainty, and that any rigorous
treatment of the planetesimal disk's evolution goes
well beyond the intended scope of this study, this model makes the simplest
possible assumption, that the planetesimal disk's dust production rate is steady
over time. This assumption is also justified by
the collisional evolution models of \cite{SC97} and \cite{K02}, who
obtain a slow erosion timescale of $\sim500$ Myrs for a Kuiper Belt orbiting at
$r\sim40$ AU. Because the timescale over which the erosion-rate varies is likely
longer than $\beta$ Pic's equilibrium
timescale ($T_{eq}\sim20$ Myrs, Section \ref{beta_pic}), the
debris disk is expected to remain in quasi-static equilibrium as it
adjusts to any slow change in the dust production rate. Consequently,
the debris-disk mass and dust production rate inferred here
are expected to be reliable.

Also note that $\beta$ Pic's dust production rate is considerable,
at least $\dot{M}_d\sim9M_\oplus/$Myr. Consequently, steady dust production
over the age of the system $t_\star$ also requires a large 
reservoir of planetesimal mass $M_p$,
with $M_p\gg \dot{M}_dt_\star\sim110$ $M_\oplus$, {\it i.e.}, the mass in
$\beta$ Pic's Kuiper Belt at $75\lesssim r\lesssim150$ AU
must be well in excess of 110 earth masses.
Alternatively, if $\beta$ Pic's Kuiper Belt does not satisfy
$M_p\gg110$ $M_\oplus$, then collisions among planetesimals
plus blowout by radiation pressure is going to grind down this
extra-solar Kuiper Belt in a few$\times t_\star\sim10$'s of Myr. In this
case the assumption of steady dust production does not hold, but it does mean that
dust production was probably
more vigorous in the past. Regardless, $\beta$ Pic's
planetesimal disk is or was very massive.

%%%%%%%%%%%%%%%%%%%%%%%%%%%%%%%%%%%
% planetesimal disk
%%%%%%%%%%%%%%%%%%%%%%%%%%%%%%%%%%%
\subsubsection{disk asymmetry}
\label{disk_asymmetry}

Many edge-on debris disks are lopsided, with one ansa substantially brighter than the
other; examples include $\beta$ Pic \citep{HLL00}, AU Mic \citep{KAG05}, and HD 15115
\citep{KFG07}. It may be that circumstellar debris disks are routinely
non-axisymmetric. Once possible explanation for a disk's asymmetry is a
recent dust-producing collision in the planetesimal disk. Recall
that the orbits of all the dust grains 
produced in a single collision have their longitudes
of periapse $\tilde{\omega}$ aligned (Section \ref{orbits}).
Because dust produced in a collision tend to loiter at their apoapse, the debris
disk will have an excess of dust at longitudes $\theta=\tilde{\omega}+\pi$.
So if that collision was vigorous enough to produce a substantial amount of
dust, then the disk will be non-axisymmetric, and may appear lopsided when
viewed edge on.

Another possible explanation for a disk's asymmetry is described in 
\cite{H09}, who showed that if the circumstellar dust is produced by an eccentric
planetesimal ring, then the smaller dust grains produced at the ring's periapse
are less bound to the star due to their higher orbital velocities there.
Consequently, there will be fewer small and marginally
bound grains in the direction of the ring's
apoapse, which will make that part of the debris disk dimmer than the periapse
side that also has an excess of small grains. This is a particularly interesting
scenario since it also implies that an extra-solar planet is likely present
in order to maintain the planetesimal ring's forced eccentricity.

Both of these scenarios will be explored in a followup study using a more
advanced version of the debris-disk model that will be generalized to handle
the collisional evolution of a non-axisymmetric debris disk.

Lastly, it should be noted that this model also ignores the planetesimals'
free eccentricity $e_f$  that is associated with the random part of their
noncircular motions. 
Equations (13-14) in \cite{TAB03} show that the planetesimals'
random motions tends to blur the simple
relationship between a dust grain's orbit and its size parameter 
(e.g., Equations \ref{a} and \ref{e}). However this effect will be small
when the planetesimals eccentricity $e_f$ is small, which in fact is a requirement
in order for planetesimals to have formed in the first place.

%%%%%%%%%%%%%%%%%%%%%%%%%%%%%%%%%%%
% Summary
%%%%%%%%%%%%%%%%%%%%%%%%%%%%%%%%%%%
\section{Summary and Conclusions}
\label{summary}

A numerical model for a circumstellar debris disk is developed and
applied to observations of $\beta$ Pictoris. The model accounts for dust
production by colliding planetesimals and dust destruction due to collisions
among grains. These rates for dust production and collisional destruction
also provide a rate equation whose solution gives
the dust abundance over time and as a
function of grain size. That solution also
provides the debris disk's grain size distribution,
which is steepened substantially
by collisions among dust grains. These calculations
also give the dust grains' collisional lifetime $T_c(R)$, which depends
on the planetesimal disk's dust production rate $\dot{M}_d$
and grain radius $R$.

Scaling laws are derived, and it is shown that the dust abundance $N$
in the debris disk varies as $N\propto\dot{M}_d^{1/2}$
once the disk has settled into collisional equilibrium, and that
dust lifetimes vary as $T_c\propto\dot{M}_d^{-1/2}$.
It is also shown that the radial drift of dust grains
due to PR drag is unimportant provided the dust production rate is sufficiently
high so that the grains' collisional lifetimes are short compared to the
drift timescale. The model also recovers the results of \cite{SC06},
who showed that the debris disk's optical depth varies as $\tau_n(r)\propto r^{-3/2}$
when in equilibrium, and that an edge-on disk's surface brightness varies
as $B(x)\propto x^{-7/2}$, where $r$ and $x$ are the radial and projected
distances from the central star. Note though that those optical depth
and surface brightness profiles will be
steeper if the disk is young and not yet in collisional equilibrium.
Alternatively, those profiles will be shallower if dust production has ceased in
the past, which might occur if planets form in (or migrate or scatter into)
the planetesimal disk and cause its dynamical depletion.

It is also shown that these quantities
are not very sensitive to the dust grains' strength $Q^\star$ when
$Q^\star<10^8$ ergs/gm. However, an edge-on disk's surface brightness profile
is quite sensitive to the dust grain's light scattering asymmetry parameter
$g$, and disks like the ones at
$\beta$ Pic and AU Mic having a `knee' in their surface
brightness indicates that the dust are very asymmetric light scatters, with
$|g|\gtrsim0.7$.

The model's principal dynamical
parameters are the planetesimal disk's radius and its dust production rate,
so a comparison to observations then yields estimates of or else limits on these
important cosmogonic quantities. For instance, fitting the model to optical
HST observations of $\beta$ Pic shows good agreement with the disk's observed
surface brightness profile when the unseen dust-producing planetesimal
disk there is quite broad, extending over $75\lesssim r \lesssim 150$ AU.
This disk's dust production rate is also quite heavy, 
$\dot{M}_d\sim9$ M$_\oplus$/Myr, if it is assumed that the dust grains
are bright like Saturn's rings ($Q_s=0.7$). In this case, the
total cross section of dust in the debris disk is
$A_d=1.9\times10^{20}$ km$^2$, and the
inferred mass there is $M_d=11$ lunar masses. Note that this mass is 
comparable to that previously inferred from submillimeter
observations of this disk \citep{HGZ98}. Also, it is unlikely
that the dust grains at $\beta$ Pic are dark ({\it i.e.}, with $Q_s<0.1$),
since that would require the planetesimal disk to grind away at a
rate that is implausible. Indeed, the mass-loss rates inferred here are
so heavy as to suggest that the $r\gtrsim75$ AU zone at
$\beta$ Pic might be a region of planetesimal destruction due to collisional grinding,
rather than a site of ongoing planet formation.

The model developed here is called {\tt ddisk},
which is an easy-to-use IDL script that others might wish to use
as they diagnose their observations of circumstellar debris disks.
This code is available for download at
{\tt http://gemelli.spacescience.org/$\sim$hahnjm/software.html}.

%%%%%%%%%%%%%%%%%%%%%%%%%%%%%%%%%%%%%%%%
% Acknowledgments 
%%%%%%%%%%%%%%%%%%%%%%%%%%%%%%%%%%%%%%%%
%\newpage
\acknowledgments

\begin{center}
  {\bf Acknowledgments}
\end{center}

Support for this work was provided by NASA through grant number HST-AR-11754.01-A
from the SPACE TELESCOPE SCIENCE INSTITUTE, which is operated by the
Association of Universities for Research in Astronomy, Inc., under NASA
contract NAS5-26555. JMH also thanks an anonymous reviewer for several helpful
suggestions.

%%%%%%%%%%%%%%%%%%%%%%%%%%%%%%%%%%%%%%%%
% References
%%%%%%%%%%%%%%%%%%%%%%%%%%%%%%%%%%%%%%%%
%\newpage
\bibliography{biblio}

\end{document}